\RequirePackage{fix-cm}
\documentclass[natbib]{svjour3}       % onecolumn (second format)
\smartqed  % flush right qed marks, e.g. at end of proof
\usepackage{mathptmx}      % use Times fonts if available on your TeX system
\usepackage{graphicx}
\usepackage{amssymb}
\usepackage{amsmath}
\usepackage{color}
\usepackage{multicol}
\usepackage{gensymb}
\usepackage{hyperref}
\hypersetup{breaklinks=true,colorlinks=true,urlcolor=blue, linkcolor=blue,  citecolor=blue,bookmarksopen=true}
\usepackage{pbox}
\usepackage{slashbox}
\usepackage[left=3cm,top=2.7cm,right=3cm,bottom=2.7cm]{geometry}
\newcommand\beq{\begin{equation}}
\newcommand\eeq{\end{equation}}
 \journalname{Space Science Reviews}
\newcommand{\nat}{{Nature}}       % Nature
\newcommand{\aj}{AJ}                   % Astronomical Journal
\newcommand{\araa}{ARA\&A}             % Annual Review of Astron and Astrophys
\newcommand{\apj}{ApJ}                 % Astrophysical Journal
\newcommand{\apjl}{ApJL}                % Astrophysical Journal, Letters
\newcommand{\apjs}{ApJS}               % Astrophysical Journal, Supplement
\newcommand{\aap}{A\&A}                % Astronomy and Astrophysics
\newcommand{\icarus}{Icarus}             % Icarus
\newcommand{\mnras}{MNRAS}             % Monthly Notices of the RAS
\newcommand{\pasp}{PASP}               % Publications of the ASP
\newcommand{\pasj}{PASJ}               % Publications of the ASJ
\begin{document}

%\title{Formation, Orbital and Internal Evolutions of Young Planetary Systems}
\title{{\huge  Formation, Orbital and Internal Evolutions of Young Planetary Systems}}
\titlerunning{Formation, Orbital and Internal Evolutions of Young Planetary Systems}

\author{Cl{\'e}ment Baruteau         \and
	    Xuening Bai			\and
	    Christoph Mordasini      \and
	    Paul Molli{\`e}re
}
\authorrunning{Baruteau, Bai, Mordasini \& Molli{\`e}re}

\institute{C. Baruteau \at
              Institut de Recherche en Astrophysique et Plan{\'e}tologie, 
              CNRS/Universit{\'e} de Toulouse, France\\
           \and
		X. Bai \at
		Harvard-Smithsonian Center for Astrophysics, Cambridge, USA\\
	\and
		C. Mordasini \at
		Physikalisches Institut, Universit{\"a}t Bern, Switzerland\\
	\and
		P. Molli{\`e}re \at
		Max-Planck-Institut f{\"u}r Astronomie, Heidelberg, Germany
	}

\date{Submitted: 22 September 2015 / Accepted: 25 April 2016}

\maketitle

\begin{abstract}
  The growing body of observational data on extrasolar planets and
  protoplanetary disks has stimulated intense research on planet
  formation and evolution in the past few years. The extremely
  diverse, sometimes unexpected physical and orbital characteristics
  of exoplanets lead to frequent updates on the mainstream scenarios
  for planet formation and evolution, but also to the exploration of
  alternative avenues. The aim of this review is to bring together
  classical pictures and new ideas on the formation, orbital and
  internal evolutions of planets, highlighting the key role of the
  protoplanetary disk in the various parts of the theory.  We begin by
  briefly reviewing the conventional mechanism of core accretion by
  the growth of planetesimals, and discuss a relatively recent model
  of core growth through the accretion of pebbles. We review the basic
  physics of planet-disk interactions, recent progress in this area,
  and discuss their role in observed planetary systems. We address the
  most important effects of planets internal evolution, like cooling
  and contraction, the mass-luminosity relation, and the bulk
  composition expressed in the mass-radius and mass-mean density
  relations.  
\end{abstract}

%==============
\section{Introduction}\label{sec:intro}
%==============
Planet formation and evolution is a fast-moving field, stimulated by
the rapid increase in the number of exoplanets and their great
diversity.  Despite the wealth of observational data on planetary
systems, including our own, it is difficult to have a general theory
for planet formation and evolution as it involves a broad range of
physical processes that happen at widely different length and time
scales. The protoplanetary disk where planets form embodies many of
the difficulties and uncertainties related to planet formation and
evolution. Protoplanetary disks are made of dust and poorly-ionized
gas governed by non-ideal magneto-hydrodynamics. The disk's structural
and turbulent properties are thus influenced by complex radiative and
chemical processes, which set the thermodynamical behavior and
ionization state of the gas. They have critical implications in every
part of planet formation and evolution: (i) they condition the location
in protoplanetary disks where dust grains grow to planetesimals and to
planets, (ii) they play a prominent role in the early orbital
evolution of low-mass planetary systems by changing the direction and
speed of planetary migration, (iii) they impact the early internal
evolution of planets by determining the amount of heat retained from
the formation process. The formation and evolution of planetary
systems is obviously affected by gravity and by the local stellar
environment. This broad range of physical processes echoes the
diversity of extrasolar planetary systems, and suggests that there is
likely no unique theory for planet formation and evolution, but
several.

Dramatic progress in understanding planet formation and evolution in
the past few years has motivated a number of recent reviews (e.g.,
\citealp{Raymond_etal14, Helled_etal14, BaruteauPP6,
  chabrierjohansen2014, baraffechabrier2014}). The aim of this chapter
is to provide an update. It brings together the formation, the orbital
and internal evolutions of young planetary systems as all three are
intimately linked. For instance, the radius and luminosity of a young
planet depends on its formation scenario, and so does its orbital
migration. The energy released by a forming planet as it cools and
contracts can change the properties of the protoplanetary disk in the
planet's vicinity and thus affect its orbital migration as much as its
growth.  This chapter thus puts an accent on, but is not restricted
to, the early stages of planetary formation and evolution, before the
dispersal of the protoplanetary disk. It begins with a review on the
growth of planetary cores by planetesimal and pebble accretions in
Sect.~\ref{sec:coregrow}.  It continues with the orbital evolution of
planets driven by planet-disk interactions, which is presented in
Sect.~\ref{sec:mig}. The internal evolution of planets follows in
Sect.~\ref{sec:theevolutionofplanets}.  Summary points are provided in
Sect.~\ref{sec:summary}.

%==============
\section{Growth of Planetesimals to Protoplanets}\label{sec:coregrow}
%==============
The journey of planet formation begins with the growth of dust grains
in protoplanetary disks and the formation of planetesimals, as
reviewed in the previous chapter by {\it Birnstiel et al.}. Due to
major uncertainties in our understanding of planetesimal formation,
most studies of the later phase of planet formation assume a
population of planetesimals are readily in place to start with, and
follow the growth of planetesimals into planetary mass objects, or
protoplanets.  If the protoplanets reach sufficiently high mass within
the lifetime of their parent protoplanetary disk, they can further
accrete gas and become the cores of giant planets (i.e., the ``core
accretion" paradigm for giant planet formation). For this reason, the
terms ``protoplanet" and ``core" are sometimes used interchangeably in
the literature.

%-------------------------------------
\subsection[]{Planetesimal Accretion}\label{ssec:plnacc}
%-------------------------------------
We begin by briefly reviewing the conventional mechanism of
planetesimal growth, where planetesimals grow by accreting other
planetesimals.  The theory has been developed and refined over
decades. Given the limited space of this chapter and the maturity of
this subject, we simply introduce the fundamental concepts of
planetesimal accretion. A pedagogical introduction can be found in the
book by \citet{Armitage10}, and a more in-depth review was given by
\citet{Goldreich_etal04}. Applications to terrestrial and extrasolar
planetary systems are summarized in the recent PPVI chapters by
\citet{Raymond_etal14} and \citet{Helled_etal14}.

% -  -  -  -  -  -  -  -  -  -  -  -  -  -  -  
\subsubsection[]{Gravitational Focusing and Runaway Growth}
% -  -  -  -  -  -  -  -  -  -  -  -  -  -  -  
Consider a swarm of planetesimals, with surface density $\Sigma_{\rm
  p}$, characteristic radial size $R$ (corresponding mass $M$), and
velocity dispersion $v_{\rm p}$. They are located around disk radius
$a$, where the disk's angular frequency is $\Omega_{\rm K}$ (equal to the Keplerian 
angular frequency). These planetesimals are thus distributed over a vertical thickness of
$H\approx v_{\rm p}/\Omega_{\rm K}$, with corresponding midplane number
density $n_{\rm p}\approx\Sigma_{\rm p}/MH\approx\Sigma_{\rm
  p}\Omega_{\rm K}/Mv_{\rm p}$.

Assuming every collision among the planetesimals leads to coagulation,
the rate of planetesimal growth is given by, within order unity,
\begin{equation}
  \frac{dM}{dt}\approx n_{\rm p}M\sigma v_{\rm rel}
  \approx\Sigma_{\rm p}\Omega_{\rm K}\sigma\ ,\label{eq:dmdt}
\end{equation}
where $v_{\rm rel}\approx\sqrt{2}v_{\rm p}$ is the typical relative
velocity of planetesimal encounters, and the corresponding collisional
cross section $\sigma$ is
\begin{equation}\label{eq:sigma}
  \sigma=\pi R^2\bigg(1+\frac{v_{\rm esc}^2}{v_{\rm rel}^2}\bigg)\ ,
\end{equation}
where $v_{\rm esc}^2=2GM/R$ is the escape velocity. The second term is
called {\it gravitational focusing}.

When the planetesimals' mass is small, or when the population of
planetesimals are ``hot" (with large velocity dispersions $v_{\rm
  p}$), we have $v_{\rm esc}\ll v_{\rm rel}$, and hence the
collisional cross section is simply geometric. Correspondingly,
$dM/dt\propto R^2\propto M^{2/3}$, or $d(\ln{M})/dt\propto
M^{-1/3}\propto R^{-1}$. This is called {\it ordered growth}. If we
rewrite the mass growth rate to growth rate in planetesimal radius, it
becomes $dR/dt\approx$const: planetesimals of all sizes grow at the
same rate.

As planetesimals become more massive, or when their relative velocity
is sufficiently low so that $v_{\rm esc}\gtrsim v_{\rm rel}$,
gravitational focusing significantly enhances the cross section. At
fixed $v_{\rm rel}$, gravitational focusing is more effective for more
massive planetesimals.  This leads to the so-called {\it runaway
  growth}, where we have $dM/dt\propto R^2\times(M/R)\propto M^{4/3}$,
or $d{\ln M}/{dt}\propto M^{1/3}\propto R$. In this regime, the growth
of larger bodies dramatically runaways over smaller bodies, and as a
result, few massive bodies stand out over the rest of the
planetesimals population, as firstly found from numerical studies
\citep{Greenberg_etal78,WetherillStewart89,KokuboIda96}.

A key requirement for runaway growth is that the velocity dispersion
$v_{\rm p}$ for the bulk of the (small) planetesimal population is
kept small.  This is a result of the balance among various dynamical
heating and cooling processes. In general, larger bodies resulting
from runaway growth are cooled by dynamical friction with small
bodies, which contain most of the mass. The small body populations are
heated during this process, but they cool by mutual collisions and gas
drag.

% -  -  -  -  -  -  -  -  -  -  -  -  -  -  -  
\subsubsection[]{Oligarchic Growth, Isolation Mass and Timescales}\label{ssec:isolation}
% -  -  -  -  -  -  -  -  -  -  -  -  -  -  -  
At late stages of the runaway growth, planetary embryos become
sufficiently large and start to interact with each other, and the
overall dynamics is dominated by these few bodies, called the {\it
  oligarchs}.  They tend to heat up the {\it neighboring}
planetesimals, increasing their velocity dispersion approximately as
$v_{\rm p}\propto M^{1/3}$. More precisely, N-body simulations suggest
that the root mean square planetesimal eccentricity $e$ and
inclination $i$ scale with embryo mass as (e.g.,
\citealp{IdaMakino93,KokuboIda00})
\begin{equation}
  \langle e^2\rangle^{1/2}\sim6R_{\rm H}/a\ ,\quad
  \langle i^2\rangle^{1/2}\sim3.5R_{\rm H}/a\ ,\label{eq:ei}
\end{equation}
where
\begin{equation}
  R_{\rm H}=\bigg(\frac{GM}{3\Omega_{\rm K}^2}\bigg)^{1/3}
  =\bigg(\frac{M}{3M_*}\bigg)^{1/3}a\label{eq:hill}
\end{equation}
is the embryo's Hill radius, and $M_*$ is the stellar mass.  This
enhancement of velocity dispersion reduces the efficiency of
gravitational focusing, and hence slows down the growth of planetary
embryos, a regime called {\it oligarchic growth}
\citep{KokuboIda98,KokuboIda00}, with growth rate scaled as
$dM/dt\propto R^2\times(M/R)/M^{2/3}\propto M^{2/3}$, or $d{\ln
  M}/dt\propto M^{-1/3}\propto R^{-1}$. Therefore, the growth mode of
the large embryos reduces to orderly. In other words, neighboring
oligarchs growth at similar rates and maintain similar masses.  On the
other hand, the collisional cross section continues to be dominated by
gravitational focusing, thus large embryos runaway over small
planetesimals. Therefore, the overall outcome of runaway growth
followed by oligarchic growth is a bi-modal distribution of an
embryo-planetesimal system.

During oligarchic growth, each planetary embryo establishes its own
domain of dominance, and maintains certain separation with neighboring
embryos. Empirically, it was found that the separation $\Delta a$ is
about $10R_{\rm H}$ \citep{KokuboIda95,KokuboIda98}. If it were
closer, planet embryos would experience close encounters, and such
strong scattering would increase their eccentricity and expand their
separation. Circularization is then achieved due to dynamical friction
from the planetesimals. Therefore, the maximum mass a planet embryo
can achieve by planetesimal accretion is limited by the amount of
planetesimals available in its {\it feeding zone}. This mass is called
the {\it isolation mass}, given by $M_{\rm iso}\approx2\pi a\Delta
a\Sigma_{\rm p}$. Assuming $\Delta a=10R_{\rm H}$ and
$M_*=M_{\bigodot}$, this mass is found to be
\begin{equation}
  M_{\rm iso}\approx50\bigg(\frac{\pi a^2\Sigma_{\rm p}}{M_*}\bigg)^{3/2}M_*
  \approx0.16\bigg(\frac{\Sigma_{\rm p}}{10\ {\rm g\ cm}^{-2}}\bigg)^{3/2}
  \bigg(\frac{a}{1\ {\rm AU}}\bigg)^3M_{\bigoplus}\ .
\end{equation}
The isolation mass mainly depends on radius and surface density of
planetesimals as $(\Sigma_{\rm p}a^2)^{3/2}$. Although the scaling
with $a$ is not well known, we generally expect $\Sigma_{\rm p}$ to
decrease with radius slower than $a^{-2}$ (e.g., $\Sigma_{\rm
  p}\propto a^{-1.5}$ in the Minimum-Mass Solar Nebula model --
hereafter MMSN), and hence $M_{\rm iso}$ increases with $a$.

We can roughly estimate the growth timescale during the oligarchic
growth phase. Using Equations~(\ref{eq:ei}), we get the typical
velocity dispersion $v_{\rm p}\sim5R_{\rm H}\Omega_{\rm K}$, and we substitute it
into Equations (\ref{eq:dmdt}) and (\ref{eq:sigma}) to obtain
\begin{equation}
\begin{split}
  \tau_{\rm grow}\approx\frac{M}{dM/dt}&\approx
  8\Omega_{\rm K}^{-1}\bigg(\frac{M_*}{\Sigma_{\rm p}a^2}\bigg)\bigg(\frac{R_{\rm H}^2}{aR}\bigg)\\
  &\approx1.2\ \bigg(\frac{M}{0.1M_{\bigoplus}}\bigg)^{1/3}
  \bigg(\frac{\Sigma_{\rm p}}{10\ {\rm g\ cm}^{-2}}\bigg)^{-1}
  \bigg(\frac{a}{1\ {\rm AU}}\bigg)^{1/2}\ {\rm Myrs}\
  .\label{eq:tgrow}
\end{split}
\end{equation}
The results above suggest that Mercury and Mars may be the leftover
embryos from oligarchic growth, and they form within a few million
years, most likely within the lifetime of the Solar nebular.  Faster
growth is possible when considering the accretion of small
planetesimals where gas drag efficiently damps their eccentricities
\citep{Rafikov04,KenyonBromley09}, the overall result is a reduction
of the pre-factor in Equation~(\ref{eq:tgrow}) but not the scaling.

Upon accreting most planetesimals from their feeding zones to achieve
the isolation mass, there is insufficient damping of random velocities
of the planetary embryos from dynamical friction
\citep{KenyonBromley06}.  This marks the end of oligarchic growth, and
the system becomes chaotic, with eccentricity growth, orbits crossing,
etc. Over secular timescales, some embryos may collide, while some may
be ejected from the system. This {\it chaotic} growth characterizes
the final stage of planet formation, and sets the mass, composition,
and orbital architecture of the planetary system.

The planetesimal accretion scenario suffers from a major difficulty
when applied to the formation of giant planet cores towards larger
separation. In the case of Jupiter and Saturn at $a\sim5-10$ AU,
reduction of $\Sigma_{\rm p}$ and increase of the dynamical time at
larger radii make the growth timescale substantially longer than in
the inner disk, as can be seen from
Equation~(\ref{eq:tgrow}). Correspondingly, the timescale to build up
a sufficiently massive core ($\sim10-15M_{\bigoplus}$,
\citealp{pollackhubickyj1996}) to trigger runaway accretion of a
gaseous envelope would well exceed the typical lifetime of a
protoplanetary disk ($\sim3$ Myrs, \citealp{Haisch_etal01}). This
issue can be alleviated when considering accretion of smaller
planetesimals \citep{Rafikov04}, enhancement of $\Sigma_{\rm p}$ by a
factor of a few beyond the snow line \citep[e.g.,][]{KK08}, and
probably resolved when type-I migration of the embryos is considered
\citep{alibertmordasini2005}.  In the latter scenario, orbital
migration via planet-disk interaction allows the embryos to sweep up
planetesimals over the course of their migration, and avoids severe
depletion of the feeding zone. Incorporating inward planet migration
into planet population synthesis models (e.g.,
\citealp{IdaLin08,mordasini2009a}), it is found that, while forming
giant planets at the location of Jupiter and Saturn is plausible, the
overall outcome depends sensitively on the strength of planetary
migration, which remains to be better understood (see
Sect.~\ref{sec:mig}).

Recently, a number of giant planets have been discovered via direct
imaging surveys to reside at large orbital separations. Notable
examples include $\beta$-Pic $b$ \citep{Lagrange_etal10} and the HR
8799 system \citep{Marois_etal10}, where orbital separation reaches up
to $\sim70$ AU.  The conventional core accretion theory via
planetesimal accretion simply fails in such cases due to the rapid
decline of planetesimal density and rapid increase of dynamical
timescale towards large orbital distances. One possible alternative to
form giant planets at large orbital separation is by the gravitational
instability (GI, \citealp{Boss97}), which likely operates in the outer
disk during the early stages of disk evolution. We will discuss GI
further in Sect.~\ref{sect:ini_cond}.

%-------------------------------------
\subsection[]{Pebble Accretion}\label{ssec:pebacc}
%-------------------------------------
Recently, a new model of planetesimal/core growth has been proposed,
where it was found that accretion of mm- to cm-sized grains, or
pebbles, can be a much more efficient mode of core growth
\citep{LambrechtsJohansen12,OrmelKlahr10,JohansenLacerda10,PeretsMurray-Clay11,MorbidelliNesvorny12}.
A large fraction of protoplanetary disks have been observed to contain
a significant fraction of dust mass in mm-sized grains
\citep{Natta_etal07,Ricci_etal10}, some also reveal the presence of cm
sized pebbles \citep{Testi_etal03,Wilner_etal05,Rodmann_etal06}.
These solids couple strongly with disk gas via aerodynamic drag, the
relative velocity can be strongly damped during gravitational
encounters, leading to substantially enhanced cross section. Because
of its relative novelty, we devote more effort discussing this new
picture of pebble accretion.

% -  -  -  -  -  -  -  -  -  -  -  -  -  -  -  
\subsubsection[]{Pebble Aerodynamics}\label{ssec:pebaero}
% -  -  -  -  -  -  -  -  -  -  -  -  -  -  -  
Due to partial pressure support, the gas in protoplanetary disks
rotates at a velocity slightly slower than the Keplerian velocity. The
velocity difference, denoted by $\Delta v_{\rm K}$, is typically a very
small fraction of the Keplerian velocity ($\sim10^{-3}-10^{-2}$
depending on location), and is commonly normalized by the local sound
speed $c_{\rm s}$. In the standard MMSN disk, it is given by $\Delta
v_{\rm K}\approx0.1(R/10{\rm AU})^{1/4}c_{\rm s}$ at disk midplane.

The pressureless dust grains, on the other hand, tend to follow
Keplerian orbits. They thus experience a headwind from the gas and
drift radially inward. The aerodynamic friction between gas and
pebbles is characterized by the stopping time $t_s$, which is commonly
normalized to the orbital time to give $\tau_s\equiv\Omega_{\rm K}
t_s$. This dimensionless stopping time is also called the Stokes
number. Strong and weak aerodynamic coupling correspond to
$\tau_s\ll1$ and $\tau_s\gg1$, respectively.  For an MMSN disk, a
spherical grain with size $s$ has a stopping time in the disk midplane
(in the Epstein regime) equal to
\begin{equation}
  \tau_s=0.14\bigg(\frac{s}{1{\rm cm}}\bigg)\bigg(\frac{R}{10{\rm AU}}\bigg)^{3/2}\ .
\end{equation}
Thus, pebble-sized grains in the outer disk are close to be marginally
coupled with the gas.  Assuming pebbles are passive in a laminar disk,
particles experience a radial drift, with sub-Keplerian rotation,
whose velocity components are given by
\begin{equation}
v_r=-\frac{2\tau_s}{1+\tau_s^2}\Delta v_{\rm K}\ ,\qquad
v_\phi-v_{\rm K}=-\frac{1}{1+\tau_s^2}\Delta v_{\rm K}\ ,\label{eq:vnsh}
\end{equation}
where $v_{\rm K}$ is the Keplerian velocity.  Radial drift is most efficient
for marginally coupled particles with $\tau_s\sim1$. In any case, the
difference between the particles velocity and the local Keplerian
velocity is still on the order of $\Delta v_{\rm K}$.

In the vertical direction, particles experience vertical gravity
$g_z=-\Omega_{\rm K}^2z$, leading to settling towards the disk
midplane. This is balanced by turbulent diffusion.  Denoting by
$D_{\rm p,z}$ the particles diffusion coefficient in the vertical
direction, particles tend to adopt a Gaussian density profile of the
form $\rho_{\rm p}=\rho_{\rm p0}e^{-z^2/2H_{\rm p}^2}$ that is
characterized by a particle scale height
\begin{equation}
H_{\rm p}\approx\sqrt{\frac{D_{\rm p,z}}{\Omega_{\rm K}\tau_s}}\ .
\end{equation}
The turbulent diffusion coefficient of the gas, $D_{\rm g}$, can be
generally written as $D_{\rm g}\sim\delta v_{\rm g}^2\times\tau_{\rm
  corr}$, where $\delta v_{\rm g}$ is the turbulent velocity of the
gas, and $\tau_{\rm corr}$ is the correlation time of the gas
turbulence, which is expected to be on the order of $\Omega_{\rm K}^{-1}$.
The particles' diffusion coefficient $D_{\rm p}$ is generally
comparable to $D_{\rm g}$ for marginally to strongly coupled particles
$\tau_s\lesssim1$ \citep{YoudinLithwick07}.  The level of turbulence
in protoplanetary disks is uncertain, but is expected to be weak
(e.g., \citealp{Bai15}). For $\delta v_{\rm g}\sim10^{-2}c_{\rm s}$,
we have $D_{\rm g}\approx D_{\rm p}\sim10^{-4}c_{\rm s}H_{\rm g}$,
with $H_{\rm g}$ the gas pressure scale height ($H_{\rm g} = c_{\rm s}
/ \Omega_{\rm K}$). The particles scale height is therefore $H_{\rm
  p}\sim0.01H_{\rm g}$ for $\tau_s=1$, and $H_{\rm p}\sim0.1H_{\rm g}$
for $\tau_s=0.01$.

% -  -  -  -  -  -  -  -  -  -  -  -  -  -  -  
\subsubsection[]{Two Regimes of Pebble Accretion}
% -  -  -  -  -  -  -  -  -  -  -  -  -  -  -  
Accurate determination of pebble accretion rates requires detailed
analysis of particle orbits as they approach the planetesimals/embryos
(e.g., \citealp{OrmelKlahr10}), and the orbits can become fairly
complex especially for loosely coupled large solids ($\tau_s$ well
above $1$). For more tightly coupled particles ($\tau_s\lesssim1$),
which are relevant in the pebble accretion scenario, simple
order-of-magnitude analysis proves to be sufficient, as we describe
here.

We consider particle trajectories as they approach the
planetesimal/core in the background gas. The gas is assumed to be
unperturbed by the core\footnote{As long as $M_c$ is much smaller than
  the ``thermal mass" $M_{\rm th}=c_{\rm s}^3/G\Omega_{\rm K}$, which is
  about the mass scale of Jupiter and is well beyond the mass scale
  for core growth.}. Therefore, if particles are very strongly coupled
to the gas, they are largely entrained by the gas without being
accreted.  More generally, particles can decouple from the gas on
timescales of their stopping time $\sim t_s$. They can be accreted if
strong gravitational deflection can be achieved within $t_s$, which
depends on the impact parameter $r$. This is the basic physics of
pebble accretion. Key in calculating the pebble accretion rate is to
estimate the maximum impact parameter ($r_a$) or the accretion radius,
within which pebbles can be accreted.

Before calculating the rate of pebble accretion, it is useful to
distinguish two regimes of pebble accretion, determined by the mass
$M_c$ of the planetesimal/core. To begin with, we can conveniently
define a Bondi radius as\footnote{Note this definition given in
  \cite{LambrechtsJohansen12} is different from the conventional
  definition of the Bondi radius, where $\Delta v_{\rm K}$ is replaced by
  the sound speed.}
\begin{equation}
  R_{\rm B}=\frac{GM_c}{\Delta v_{\rm K}^2}\ .
\end{equation}
This definition of $R_{\rm B}$ marks the length scale at which pebbles can
be significantly deflected by the planetesimal/core by two-body
interactions.  Additionally, this deflection can not be effective
beyond the Hill radius $R_{\rm H}$ (Equation~\ref{eq:hill}), where
three-body effects become important and the core loses gravitational
dominance.

By equating $R_{\rm B}$ and $R_{\rm H}$, the {\it transition mass} can be defined as
\begin{equation}
  M_t=\frac{\Delta v_{\rm K}^3}{G\Omega_{\rm K}}\approx
  0.16 M_\oplus\bigg(\frac{R}{30\rm AU}\bigg)^{3/4}
  \bigg(\frac{\Delta v_{\rm K}}{0.1c_{\rm s}}\bigg)^3\ ,
\end{equation}
where in the second equality we use the MMSN scaling for the sound
speed.  Note that $R_{\rm B}\propto M_c$ while $R_{\rm H}\propto M_c^{1/3}$, thus
$(R_{\rm B}/R_{\rm H})\approx(M_c/M_t)^{2/3}$. Therefore, the distance from the
planetesimal/core within which its gravitational influence is
effective is given by $R_{\rm B}$ for low-mass cores, and by $R_{\rm H}$ for
high-mass cores. Correspondingly, the transition mass defined above
separates the two regimes of pebble accretion, which are termed {\it
  drift regime} for $M_c<M_t$, and {\it Hill regime} for $M_c>M_t$.
Estimates of the accretion radius $r_a$ differ between both regimes,
as we address below.

% -  -  -  -  -  -  -  -  -  -  -  -  -  -  -  
\subsubsection[]{Drift Regime}\label{ssec:drift}
% -  -  -  -  -  -  -  -  -  -  -  -  -  -  -  
In the drift regime where $M_c<M_t$ (or, equivalently, the Bondi
radius $R_{\rm B}<R_{\rm H}$), the accretion radius $r_a\leq R_{\rm B}$. Within this
radius, particles approach the core with relative velocity $\Delta
v\sim\Delta v_{\rm K}$. The timescale for a particle to pass by the core is
\begin{equation}
  t_B\approx\frac{R_{\rm B}}{\Delta v_{\rm K}}
  =\frac{1}{\Omega_{\rm K}}\frac{M_c}{M_t}\ .
\end{equation}
Particles with stopping time $t_s\approx t_B$ decouple from the gas
during the encounter, and are expected to spiral into the core once
their impact parameter is within $R_{\rm B}$. Therefore, the accretion
radius $r_a\approx R_{\rm B}$ for $t_s\approx t_B$.

For more strongly coupled particles with $t_s<t_B$, they can only be
captured when their impact parameter is smaller than $R_{\rm B}$, so that
they experience stronger gravity with shorter gravitational deflection
time (and hence they can decouple from the gas and spiral
in). Otherwise, they will be entrained by the gas flow without being
accreted.  The gravitational deflection time for given impact
parameter $r$ is
\begin{equation}
  t_g(r)\approx\frac{\Delta v_{\rm K}}{GM_c/r^2}=\bigg(\frac{r}{R_{\rm B}}\bigg)^2t_B\ .
\end{equation}
By equating $t_g(r)$ with $t_s$, the accretion radius $r_a$ can be
obtained as
\begin{equation}
  r_a\approx\bigg(\frac{t_s}{t_B}\bigg)^{1/2}R_{\rm B}
  \approx\tau_s^{1/2}\bigg(\frac{M_c}{M_t}\bigg)^{1/6}R_{\rm H}\ .\label{eq:radrift}
\end{equation}
This radius can equivalently be understood as the radius at which the
gravitational acceleration is equal to the acceleration due to gas
drag (also termed as the wind-shearing radius, \citealp{PeretsMurray-Clay11}).

For more weakly coupled particles with $t_s>t_B$, they behave more
similarly to the gas-free scenario, where particles follow hyperbolic
orbits as they encounter the core. Instead of the Bondi radius, it is
the physical size of the core (much smaller) that determines whether
accretion takes place. This results in a rapid decline of $r_a$ as
$t_s$ increases beyond $t_B$, and it becomes negligible compared with
the $t_s\sim t_B$ case.

We note that in the drift regime, $\Omega_{\rm K} t_B\approx M_c/M_t$.
Therefore, efficient accretion of dust grains in the drift regime
always takes place for particles with $\tau_s<1$. The smaller the core
mass, the smaller the optimal particle size.

% -  -  -  -  -  -  -  -  -  -  -  -  -  -  -  
\subsubsection[]{Hill Regime}\label{ssec:hill}
% -  -  -  -  -  -  -  -  -  -  -  -  -  -  -  
In the Hill regime where $M_c>M_t$ (or, equivalently, the Bondi radius
$R_{\rm B}>R_{\rm H}$), we expect the accretion radius $r_a\leq R_{\rm H}$. Within this
radius, the relative velocity with which particles approach the core
can be dominated by the Keplerian shear, $\Delta v_{\rm
  sh}(r)\equiv(3/2)\Omega_{\rm K}r$.  To see this, note that
\begin{equation}
\frac{\Delta v_{\rm sh}(r)}{\Delta v_{\rm K}}
=\frac{3}{2}\bigg(\frac{M_c}{M_t}\bigg)\bigg(\frac{r}{R_{\rm B}}\bigg)\ .
\end{equation}
Therefore, the shear velocity dominates for impact parameters
$r\gtrsim R_{\rm B}$. For a given impact parameter $r$, the relative
velocity is $\Delta v(r)\approx\max{[\Delta v_{\rm K}, \Delta v_{\rm
    sh}(r)]}$.

Following the same spirit as before, we write the gravitational
deflection time for particles as
\begin{equation}
  t_g(r)=\frac{\Delta v}{GM_c/r^2}\approx\frac{\Omega_{\rm K}r}{GM_c/r^2}
  \approx\frac{1}{\Omega_{\rm K}}\bigg(\frac{r}{R_{\rm H}}\bigg)^3\ ,
\end{equation}
where we have taken $\Delta v\approx\Omega_{\rm K}r$. Efficient pebble
accretion requires $t_g\lesssim t_s$, which leads to an accretion
radius of
\begin{equation}
  r_a\approx(\Omega_{\rm K}t_s)^{1/3}R_{\rm H}=\tau_s^{1/3}R_{\rm H}\ .\label{eq:rahill}
\end{equation}
This result again holds\footnote{By integrating individual particle
  trajectories, $r_a$ is found to be about $R_{\rm H}$ for particles with
  $\tau_s\sim0.1$ \citep{LambrechtsJohansen12}, thus
  $r_a\approx(\tau_s/0.1)^{1/3}R_{\rm H}$ is probably more accurate for
  $\tau_s\lesssim0.1$, which still agrees with the order-of-magnitude
  derivation within a factor of order unity.}  for particles with
$\tau_s\lesssim1$.Particles with lager stopping times would undergo
3-body scattering whose orbits can become fairly chaotic. While gas
drag still enhances particle accretion, accretion is less efficient
with increasing $\tau_s$, and more over, we do not expect a large
population of particles with stopping times far exceeding $1$ to be
present in protoplanetary disks.

The above scaling needs to be revised when $r_a<R_{\rm B}$, at which point
we should have taken $\Delta v=\Delta v_{\rm K}$. The revised $r_a$ then
becomes the expression in Equation~(\ref{eq:radrift}).  Therefore, the
overall result is
\begin{equation}
r_a\approx\min{\bigg[\tau_s^{1/2}\bigg(\frac{M_c}{M_t}\bigg)^{1/6},
\tau_s^{1/3}}\bigg]R_{\rm H}\ .\label{eq:rahill2}
\end{equation}

The fact that the accretion radius approaches the Hill radius in the
Hill regime makes pebble accretion an extremely efficient mechanism
for planetesimal/core growth. Additionally, the weak dependence of the
accretion radius on $\tau_s$ in Equation (\ref{eq:rahill2}) indicates
efficient accretion takes place for a wide range of particle sizes
(say $\tau_s=0.01-1$).

% -  -  -  -  -  -  -  -  -  -  -  -  -  -  -  
\subsubsection[]{Timescale for Core Growth}
% -  -  -  -  -  -  -  -  -  -  -  -  -  -  -  
The rate of pebble accretion is determined by the accretion radius
$r_a$, and the radial flux of pebbles approaching the core. To
calculate it, we need to distinguish the 2D and 3D cases, depending on
whether the particles' scale height $H_{\rm p}$ is larger or smaller
than $r_a$. In the following, we denote by $\Sigma_{\rm p}$ the
surface density of the pebbles.

When $H_{\rm p}<r_a$, we are in the 2D regime where particles settle
strongly to the midplane so that the entire particle column
approaching the core can be accreted, with accretion rate
\begin{equation}
\dot{M}_{2D}\approx2\Sigma_{\rm p}r_a\Delta v\ .
\end{equation}
When $H_{\rm p}\gtrsim r_a$, we are in the 3D regime where particles
are suspended, and only those close to the midplane with $|z|\lesssim
r_a$ can be accreted. The accretion rate is given by
\begin{equation}
\dot{M}_{3D}\approx\frac{\Sigma_{\rm p}}{H_{\rm p}}\pi r_a^2\Delta v\ .
\end{equation}
In both cases, we have $\Delta v\approx\max{[\Delta v_{\rm K}, \Delta
  v_{\rm sh}(r_a)]}$.

Now we can discuss the planetesimal/core growth rate due to pebble
accretion. While multiple combinations of 2D/3D and drift/Hill regimes
are possible, here we simply discuss two most representative cases.

In the early stage of planet formation when the planetesimal/core is
small, pebble accretion is in the drift regime with $R_{\rm B}<R_{\rm H}$, and
likely proceeds in 3D. In that case, we get
\begin{equation}
\dot{M}_c\approx\pi\frac{\Sigma_{\rm p}}{H_{\rm p}}
\bigg(\frac{t_s}{t_B}R_{\rm B}^2\bigg)\Delta v_{\rm K}\ .
\end{equation}
In the optimistic case where the contribution from $t_s/t_B\sim1$
particles is considered ($t_s$ increases with $M_c$), growth can
achieve super-runaway with $\dot{M}_c\sim M_c^2$, although the base
rate (i.e., proportional coefficient) can be very low.

Towards later stages of planet formation, when the core has grown
substantially across the transition mass, accretion proceeds in the
Hill regime and can operate in 2D if turbulence is not too strong. The
corresponding accretion rate is
\begin{equation}
\dot{M}_c\approx2\Sigma_{\rm p}\tau_s^{2/3}R_{\rm H}^2\Omega_{\rm K}\ .
\label{Mdot_Hill_3D}
\end{equation}
In terms of scaling, we obtain $\dot{M}_c\propto M_c^{2/3}$, which is
the same rate as with oligarchic growth. However, the proportional
coefficient is much larger than in the planetesimal accretion
scenario, because of the greatly enhanced cross section of the order
$R^2_{\rm H}$.

% FFFFFFFFFFFFF
\begin{figure}
\centering
\includegraphics[width=0.8\hsize]{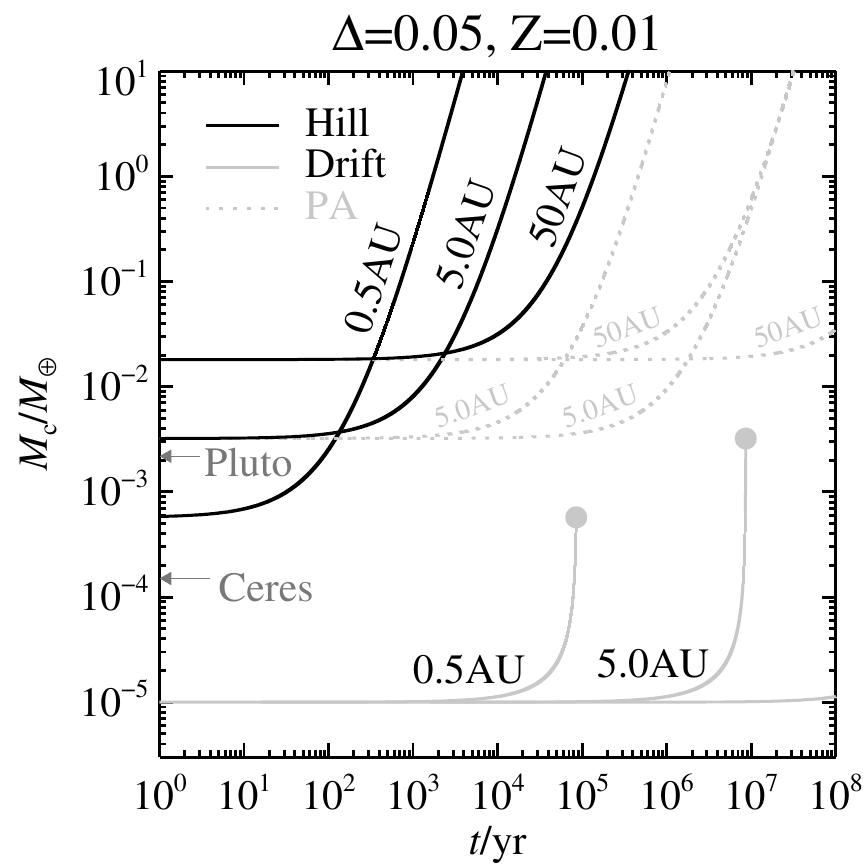}
\caption{Core growth as a function of time for various orbital
  distances for the pebble accretion scenario in the Hill and drift
  regimes, as well as the conventional planetesimal accretion (denoted
  by ``PA"), taken from \cite{LambrechtsJohansen12}. This calculation
  assumes parameter $\Delta\equiv\Delta v_{\rm K}/c_{\rm s}=0.05$, and
  dust-to-gas mass ratio $Z=0.01$ in a minimum-mass solar nebular
  disk.}
\label{fig:PA}
\end{figure}
% FFFFFFFFFFFFF

To evaluate the core growth timescale via pebble accretion (in the
Hill regime), we first rewrite Equation~(\ref{Mdot_Hill_3D}) into the
more intuitive form
\begin{equation}
  \frac{1}{M_c}\frac{dM_c}{dt}\approx\frac{\Sigma_{\rm p}r^2}{M_*}
  \bigg(\frac{M_*}{M_c}\bigg)^{1/3}\Omega_{\rm K}\ ,
\end{equation}
where we consider the most optimistic case that all particles entering
the Hill sphere get accreted. This roughly corresponds to the case
where most dust particles have stopping time $\tau_s\sim0.1-1$ (and 2D
accretion), which is probably roughly satisfied in the outer regions
of protoplanetary disks. We note that the growth rate is independent
of $\Delta v_{\rm K}$ (but the transition mass does). Assuming a Solar mass
star, the growth time is
\begin{equation}
  t_{\rm grow}\approx2.2\times10^3\bigg(\frac{2\pi}{\Omega_{\rm K}}\bigg)
  \bigg(\frac{0.005M_{\bigodot}}{\Sigma_{\rm p}r^2}\bigg)
  \bigg(\frac{M_c}{M_{\bigoplus}}\bigg)^{1/3}\ .
\end{equation}
Assuming a typical disk mass of $\sim0.01M_{\bigodot}$, and that the
planet is formed half-way, then for an earth-mass core the growth time
is about 2000 orbits, which is only $\sim3.6\times10^5$ years at 30
AU.

Figure~\ref{fig:PA} shows a simple model for the time evolution of the
core's mass resulting from the conventional planetesimal accretion as
well as from pebble accretion, taken from
\citet{LambrechtsJohansen12}.  The calculations correspond to the
optimistic 2D accretion case with maximum accretion rate. The general
conclusions are that pebble accretion in the Hill regime is extremely
efficient, especially at large orbital distances. On the other hand,
in the drift regime, pebble accretion is slow and core growth
generally proceeds more efficiently via planetesimal accretion.

% -  -  -  -  -  -  -  -  -  -  -  -  -  -  -  
\subsubsection[]{Global Models and Uncertainties}\label{ssec:glob}
% -  -  -  -  -  -  -  -  -  -  -  -  -  -  -  
The general formulation of pebble accretion has recently been
incorporated in global models of planet formation
\citep{Chambers14,KretkeLevison14,LambrechtsJohansen14,Bitsch15}.  In
general, these models confirm that pebble accretion is indeed very
efficient and can well account for the formation of the cores of not
only the gas and ice giant planets in the Solar system, but also the
extrasolar giant planets at large separations. The growth might be
fast enough to avoid large-scale type I planetary migration (see
Sect.~\ref{ssec:type1}).  Additionally, a large fraction ($\sim50\%$)
of pebbles undergoing radial drift can be intercepted by the
embryos/planetesimals, leading to very efficient conversion of solids
into planets \citep{LambrechtsJohansen14}. On the other hand,
formation of gas giants is less favored in systems with relatively
small disk masses and/or metallicities, consistent with the
metallicity trend in the occurrence rate of giant exoplanets
\citep{FischerValenti05,Bitsch15,Johnson_etal10}.

Applying to the Solar system, criticisms arise concerning the fact
that it may be too efficient: the system tends to form a large number
of planetary embryos inside-out, which would lead to violent dynamical
interactions and can hardly be made consistent with the planetary
architecture of the inner Solar system \citep{KretkeLevison14}.  On
the other hand, \citet{Morbidelli_etal15} pointed out that pebble
accretion is likely very inefficient within the snow line due to
reduced grain size, which readily explains the dichotomy of low-mass
terrestrial planets in the inner Solar system and gas giants in the
outer parts.

At the present stage, studies from different groups have adopted
different methodologies in treating various physical processes during
planetesimal/core growth, and inevitably involve simplified
assumptions in one way or another. As also noticed in these works,
major uncertainties exist in these modeling efforts, which include:
1). The size distribution and evolution of pebbles, which were either
treated as the end product of grain growth process, or as fragments
from planetesimal collision. 2). The initial mass function of
planetesimals, the starting point of planetesimal growth.  3). Global
disk structure and evolution, and level of turbulence, particularly
relevant for the dynamics and transport of pebbles.  None of these
factors are well understood, and they are interrelated with each
other.

The key to unravel these uncertainties probably lies in better
understanding the internal structure and evolution of protoplanetary
disks from first principle. Recent local 3D MHD simulations including
realistic non-ideal MHD physics have suggested a paradigm shift from
the conventional viscous evolution picture towards a magnetized disk
wind driven evolution of the inner disk \citep{BaiStone13b,Gressel15},
and a weakly turbulent outer disk with substructures due to magnetic
flux concentration \citep{BaiStone14,Bai15}. Extension of these local
studies to global simulations is essential and will likely help
establish a most realistic picture of protoplanetary disks. Such a
picture will provide essential input for both grain growth
calculations and necessary conditions for planetesimal formation (see
Chapter by {\it Birnstiel et al.}), and altogether, for modeling the
entire processes of planet formation.

%==============
\section{Orbital Evolution of Planets in their Protoplanetary Disk}\label{sec:mig}
%==============
As planets form in their protoplanetary disk, the gravitational
interaction between the disk and the planets change the semi-major
axis, eccentricity and inclination of the planets. Planet-disk
interactions usually damp eccentricities and inclinations efficiently,
but as we will see throughout this section, they can either decrease
or increase the planets semi-major axes, an effect usually termed
planetary migration or disk migration. Planet-planet interactions tend
to pump eccentricities and inclinations, and can significantly change
semi-major axes during the disk's lifetime, for example through
scattering events due to the disk migration of two or more planets.
After the protoplanetary disk is cleared by photo-evaporation, by
typically 1 to 10 Myrs, the orbital evolution of planets can continue
through interaction with the central star (via tides or stellar
evolution for close-in planets), with distant stars (e.g., via Kozai
cycles), with a disk of remnant planetesimals (debris disk), or via
planet-planet interactions.

As this chapter is about the early evolution of planetary systems, we
will focus on the orbital evolution of planets driven by planet-disk
interactions. The long-term evolution of planetary systems, after
dispersal of the protoplanetary disk, will be presented in the Chapter
by {\it Zhou et al.} The aim of this section is to give a simple,
practical overview of the physical processes that drive planet-disk
interactions. The case of low-mass planets (typically Earth-mass
planets) is described in Sect.~\ref{ssec:mig_low}, that of massive
planets (typically Jovian planets) follows in
Sect.~\ref{ssec:mig_high}.  We then discuss in
Sect.~\ref{ssec:mig_obs} what models of planet-disk interactions can
and cannot explain about hot Jupiters (Sect.~\ref{ssec:obliq}), and
about the many super-Earths in multi-planet systems discovered by {\it
  Kepler} (Sect.~\ref{ssec:multi}). More in-depth reviews of
planet-disk interactions can be found in \cite{KleyNelson12},
\cite{BM13} and \cite{BaruteauPP6}.

%-------------------------------------
\subsection[]{Disk Migration of Low-Mass Planets}
\label{ssec:mig_low}
\begin{figure}
\centering
\includegraphics[width=\hsize]{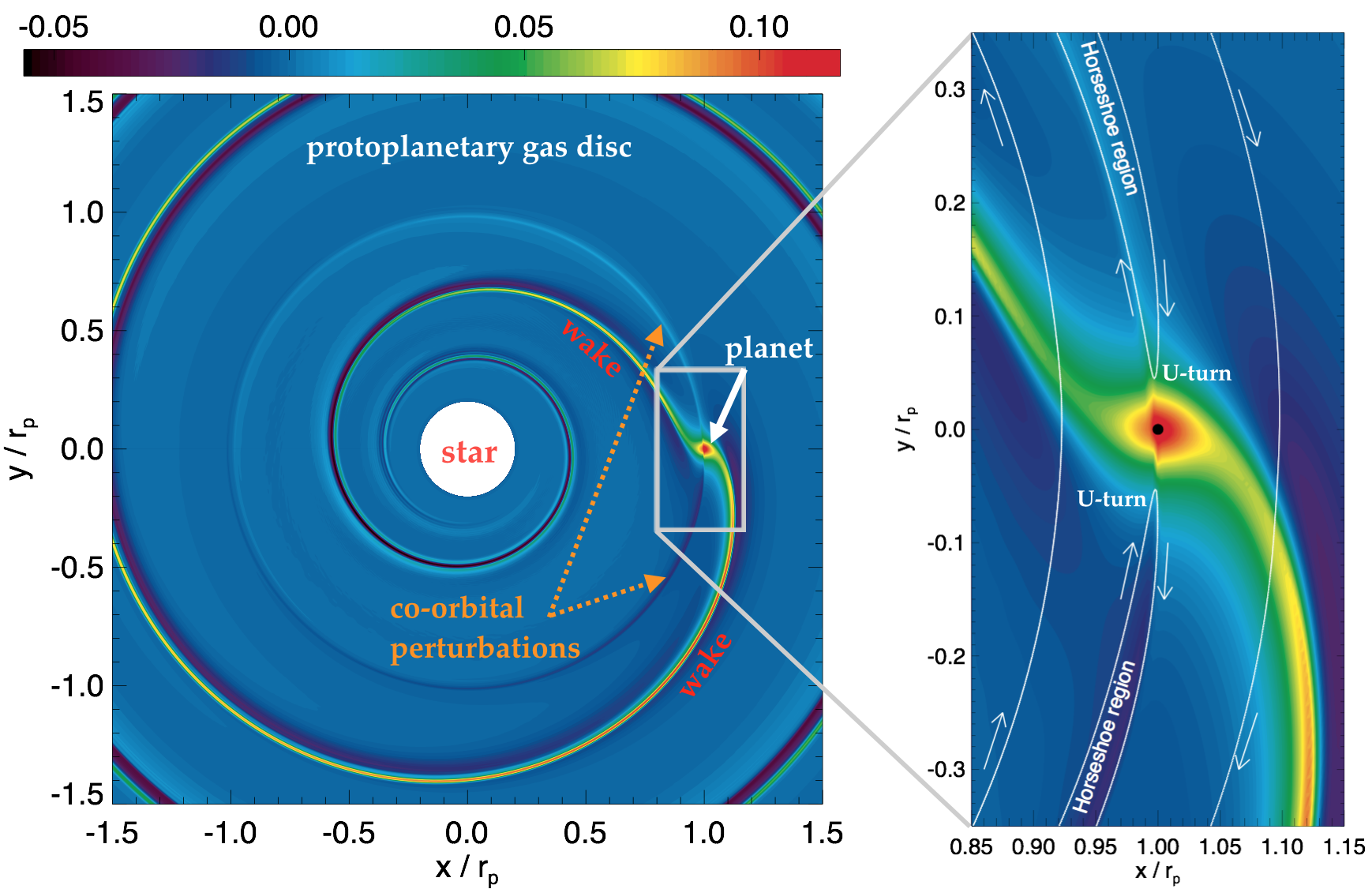}
\caption{Relative perturbation of the gas surface density of a
  protoplanetary disk perturbed by a 5 Earth-mass planet. The planet's
  location is spotted by the white arrow in the left panel. The planet
  generates a one-armed spiral density wave -- the wake -- that
  propagates throughout the disk, as well as co-orbital density
  perturbations within the planet's horseshoe region.  Gas
  trajectories with respect to the planet are depicted by white curves
  and arrows in the right panel. Figure adapted from
  \cite{BaruteauPP6}.  }
\label{fig:CB1}
\end{figure}

% -------------------------------------
Since planetary migration is driven by the gravitational interaction
between a planet and its parent protoplanetary disk, the basics of
planetary migration can be drawn through inspection of the gas density
perturbation caused by a protoplanet. Figure~\ref{fig:CB1} shows the
relative perturbation of the gas surface density of a protoplanetary
disk where a 5 Earth-mass planet forms.  The figure is taken from a 2D
hydrodynamical simulation of planet-disk interactions for a
non-magnetized radiative disk model. As can be seen in the figure, the
planet induces two kinds of density perturbations in the disk: (i)
spiral density waves, called the planet's wakes, which propagate
throughout the disk, and (ii) density perturbations very close to the
planet's orbital radius, called co-orbital density perturbations,
which are confined in the planet's horseshoe region (see right panel),
where gas describes horseshoe trajectories relative to the planet.
% FFFFFFFFFFFFF
% FFFFFFFFFFFFF

The basic idea underlying planet-disk interactions is the law of
action-reaction: the disk reacts to the planet's gravity by exerting a
gravitational force on the planet which changes the planet's
semi-major axis, eccentricity and inclination. Considering the
simplest case of a planet on a circular and coplanar orbit (with zero
eccentricity and inclination), the planet's angular momentum is
$J_{\rm p} = M_{\rm p} \sqrt{GM_{\star}r_{\rm p}}$, where $M_{\rm p}$
is the mass of the planet, $M_{\star}$ that of the star, $G$ is the
gravitational constant, and $r_{\rm p}$ is the distance between the
star and the planet. Denoting by $\Gamma$ the torque exerted by the
disk on the planet, and further assuming the planet mass is
stationary, $\Gamma = dJ_{\rm p} / dt$ implies $dr_{\rm p} / dt =
\Gamma \times 2r_{\rm p}/J_{\rm p}$. This relation simply means that,
in order to know the direction and speed of planetary migration, the
sign and magnitude of the disk torque need to be determined. This is
why most numerical studies of planet-disk interactions take planets on
fixed orbits and calculate the disk torque on the planet. However, as
we will see later, the disk torque can actually be a function of the
migration rate and migration history, with the consequence that, under
some circumstances, the migration can runaway. We briefly outline
below the different components of the disk torque.

% -  -  -  -  -  -  -  -  -  -  -  -  -  -  -  
\subsubsection[]{Wake Torque}\label{ssec:wake}
% -  -  -  -  -  -  -  -  -  -  -  -  -  -  -  
First, there is the torque due to the inner wake, which is the wake
that propagates between the planet and the star. Most of the inner
wake's torque comes from the excess of gas just in front of the planet
in the azimuthal direction, and which is distant from the planet by a
few pressure scale heights at most. This gas excess exerts a
gravitational force on the planet whose azimuthal component
$F_{\varphi}$ is positive, which corresponds to a torque $\sim r_{\rm
  p} F_{\varphi}$ on the planet that is therefore positive. Similarly,
the outer wake's torque is mostly due to the excess of gas immediately
behind the planet in the azimuthal direction.  That local gas excess
now has $F_{\varphi} < 0$ and thus exerts a negative torque on the
planet. The total wake torque then has two opposite contributions: the
(positive) torque due to the inner wake that tends to move the planet
outwards, and the (negative) torque due to the outer wake that tends
to move the planet inwards. Inspection of the inset panel in the
right-hand side of Figure~\ref{fig:CB1} shows that the gas density
perturbation in the outer wake is slightly larger and closer to the
planet, which indicates that the outer wake's torque is stronger. The
total wake torque is therefore negative in that case and favors inward
migration. While this is indeed the general expectation, the reader
should be aware that this result depends on the radial gradients of
the gas temperature and density near the planet's location, as they
affect the location of the wakes relative to the planet as well as
their density enhancement. These dependencies can be found by solving
the linear perturbation equations numerically, either in 2D
\citep{KP93,w97,pbck10,Masset11} or in 3D \citep{tanaka2002}. The
reader is referred to Sect.~2.1.1 of \cite{BaruteauPP6} for more
details on the wake torque and its expression as a function of disk
gradients.

% -  -  -  -  -  -  -  -  -  -  -  -  -  -  -  
\subsubsection[]{Corotation Torque}\label{ssec:corotation}
% -  -  -  -  -  -  -  -  -  -  -  -  -  -  -  
The disk gas in the planet's horseshoe region follows horseshoe
trajectories as seen from the planet. At its closest approach to the
planet, the gas undergoes a gravitational kick from the planet.  The
gas just behind the planet in the azimuthal direction and inside the
planet's orbital radius moves radially outward by taking angular
momentum from the planet. This gas takes outward U-turns relative to
the planet and exerts a negative torque on the planet.  Meanwhile, gas
just in front on the planet in the azimuthal direction and outside the
planet's orbital radius moves radially inward by giving angular
momentum to the planet. That gas embarks on inward U-turns and exerts
a positive torque on the planet. Much like the wake torque, the torque
exerted by the horseshoe region, which is called the corotation torque
or horseshoe drag, has two opposite contributions. The sign and
magnitude of the corotation torque then depend on the angular
momentum difference between gas doing inward and outward U-turns.

Complexity arises as the angular momentum of the gas in the planet's
horseshoe region evolves in time due to the advection-diffusion of two
hydrodynamical quantities: (i) the gas specific vorticity, sometimes
called vortensity, which in 2D is the vertical component of the
velocity curl (vorticity) divided by the surface density, and (ii) the
gas specific entropy, which in 2D is basically the quantity $P
\Sigma^{-\gamma}$ with $P$ the thermal pressure, $\Sigma$ the surface
density and $\gamma$ the adiabatic index. Advection of vortensity and
specific entropy along horseshoe streamlines implies that the sign and
magnitude of the corotation torque depend on the density and
temperature gradients across the horseshoe region. Diffusion of both
quantities implies that the corotation torque also depends on the
nature and efficiency of turbulent diffusion mechanisms taking place
near or inside the horseshoe region, which for low-mass planets is a
small fraction of the disk's pressure scale height. Modelling of
turbulence by a viscosity and a thermal diffusivity shows that the
corotation torque is very sensitive to both diffusion
parameters. Hence the importance of better understanding how turbulent
transport of angular momentum operates in regions of planet formation
\citep[e.g.,][]{Lesur14,Bai15,Gressel15}. 2D and 3D models of viscous
disks show that the corotation torque is usually positive and
therefore favors outward migration \citep[e.g.,][]{pm06,bm08a, pp08,
  KC08, KBK09, MC10, pbck10, PBK11,Lega14}. The reader is referred to
Sect.~2.1.2 of \cite{BaruteauPP6} for more details on the corotation
torque. Its expression in 2D as a function of disk gradients,
viscosity and thermal diffusivity can be found in \cite{MC10} or
\cite{PBK11}.  A derivation of the corotation torque in 3D globally
isothermal disk models has been recently presented by
\cite{MassetBenitez16}.  Dynamical corotation torques, which arise as
the planet drifts relative to the disk, will be described in
Sect.~\ref{ssec:mig_progress}.

% -  -  -  -  -  -  -  -  -  -  -  -  -  -  -  
\subsubsection[]{Type I Migration}\label{ssec:type1}
\begin{figure}
\centering
\includegraphics[width=0.9\hsize]{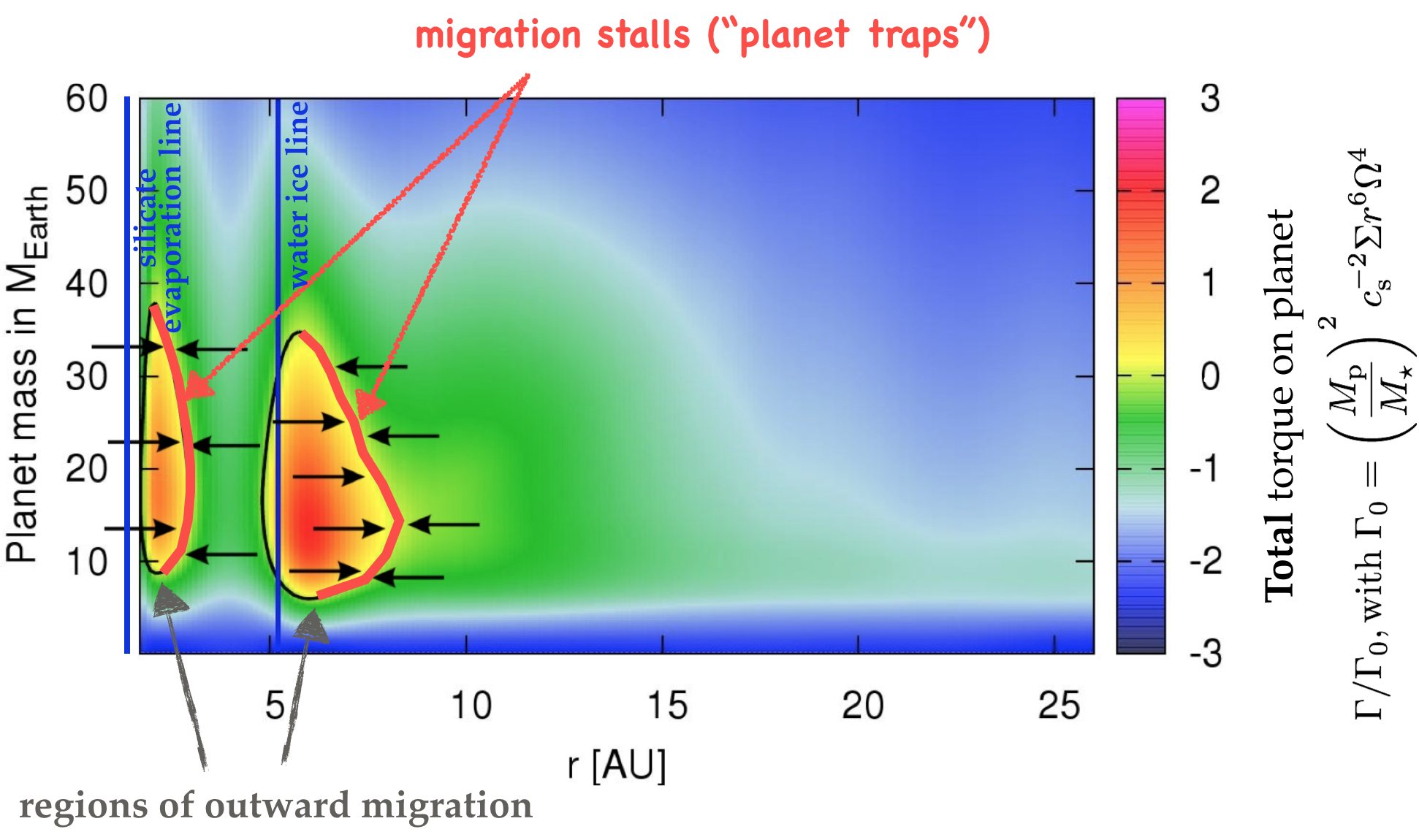}
\caption{Type I migration torque on planets of different masses
  (y-axis) and orbital radii (x-axis). Horizontal arrows show the
  direction of migration. Figure adapted from \cite{Bitsch13}, where
  the disk model sets the water ice line at $r \approx 5$ AU, and the
  evaporation line of silicates at $r \approx 0.7$ AU.}
\label{fig:CB2}
\end{figure}

% -  -  -  -  -  -  -  -  -  -  -  -  -  -  -  
The migration of low-mass planets (typically up to Neptune's mass),
known as type I migration, is driven by the sum of the wake torque
(Sect.~\ref{ssec:wake}) and the corotation torque
(Sect.~\ref{ssec:corotation}). The former is usually negative, the
latter usually positive. Once more, we have two competing effects, and
the balance between them is very sensitive to the disk model. This is
illustrated in Figure~\ref{fig:CB2}, which displays the type I
migration torque with varying planet mass and planet's orbital
radius. In this disk model taken from \cite{Bitsch13}, opacity
transitions near the evaporation line of silicates (at $r \approx 0.7$
AU) and the water ice line (at $r \approx 5$ AU) boost the corotation
torque beyond these transitions, where a broad range of planet masses
undergoes outward migration. Note that type I migrating cores converge
towards the outer edge of these regions of outward migration; these
locations are known as ``planet traps".
% FFFFFFFFFFFFF
% FFFFFFFFFFFFF

The torque value in Figure~\ref{fig:CB2} can be used to estimate the
type I migration timescale. The factor $\Gamma_0$ by which the torque
is normalized in the color bar sets indeed a typical timescale for
type I migration, $\tau_0$, given by $\tau_0 = \Omega_{\rm K} M_{\rm p}
r^2_{\rm p} / 2\Gamma_0$, and which can be recast as
\begin{equation}
\tau_0 \approx 1.3{\,\rm Myr}
\times \bigg( \frac{h}{0.05} \bigg)^2
\bigg( \frac{\Sigma}{200\,{\rm g\,cm}^{-2}} \bigg)^{-1}
\bigg( \frac{M_{\star}}{M_{\odot}} \bigg)^{3/2}
\bigg( \frac{r_{\rm p}}{5\,{\rm AU}} \bigg)^{-1/2}
\bigg( \frac{M_{\rm p}}{M_{\oplus}} \bigg)^{-1},
\label{eq:tau_0}
\end{equation}
with $h=c_{\rm s}/v_{\rm K}$ the disk's aspect ratio, and where all
gas quantities are to be calculated at the planet's orbital radius.
Above estimate shows that an Earth-mass planet located at 5 AU from a
Solar-mass star, with local gas surface density of 200 g cm$^{-2}$ and
disk's aspect ratio of 5\%, and which experiences a total disk torque
$\Gamma \approx -2\Gamma_0$ (see Figure~\ref{fig:CB2}), would migrate
inwards on a characteristic timescale of $\sim 0.7$ Myr. This should
be compared with the typical growth timescale of the planet. In the
Hill regime of pebble accretion, the growth timescale of such a planet
could be as short as $\sim 2\times 10^{4}$ yr (see the black bold
curves in Figure~\ref{fig:PA}), which suggests that Earth-mass planets
could grow fast enough to avoid significant inward migration and reach
planetary traps (the red bold curves in
Figure~\ref{fig:CB2}). Planetesimal accretion, however, would result
in a growth timescale at least comparable to the aforementioned
migration timescale (see the dashed curves in Figure~\ref{fig:PA}), in
which case significant inward migration would occur.

There are several points to bear in mind about
Figure~\ref{fig:CB2}. First is that this disk model assumes a
turbulent disk viscosity associated with a constant alpha viscous
parameter $\alpha \sim {\rm a\,few} \times 10^{-3}$. Different values
of $\alpha$, or perhaps more realistically a radial profile of
$\alpha$ inspired from simulations of MHD turbulent disks, would
result in different locations for regions of outward migration
\citep{Bitsch14}. Second is that this picture is time-dependent: as
the disk evolves, it loses mass and cools with the consequence that
regions of outward migration progressively shift towards smaller
orbital radii and smaller planet masses \citep{Bitsch15mig}.
 
% -  -  -  -  -  -  -  -  -  -  -  -  -  -  -  
\subsubsection[]{Some Recent Progress on Type I Migration}\label{ssec:mig_progress}
% -  -  -  -  -  -  -  -  -  -  -  -  -  -  -  
The predicted rapid inward migration of planets of few Earth masses
has stimulated much work on type I migration. A selection of recent
works on type I migration is detailed below.

{\bf (i) Disk's magnetic field -- } The recent work by \cite{guilet13}
has shown that there is an additional corotation torque on low-mass
planets when the disk has a weak toroidal magnetic field. This new
torque comes about because of advection-diffusion of the magnetic flux
inside the planet's horseshoe region. The horseshoe trajectories of
the gas cause an accumulation of the magnetic flux contained in the
horseshoe region downstream of the U-turns. The magnetic pressure
increases locally, and the thermal pressure decreases to maintain a
mechanical equilibrium with the surrounding gas. This, in turn,
decreases the local gas surface density if the disk's temperature is
stationary. An azimuthal asymmetry of the horseshoe U-turns relative
to the planet, which is related to the radial pressure gradient, makes
this new corotation torque generally positive and favors outward
migration. Its magnitude depends sensitively on the strength of the
magnetic field and the magnetic resistivity \citep{guilet13}. We point
out that the surface density perturbation obtained in the 2D laminar
disk models of \cite{guilet13}, where the effects of turbulence are
modelled by a viscosity and a magnetic resistivity, is in excellent
agreement with the time-averaged density perturbation found in the 3D
MHD turbulent simulations of \cite{bfnm11}, which were for Saturn-mass
planets in high aspect ratio disks (see the comparison in Figure~3 of
\citealp{BaruteauPP6}). More work is needed in this area, in
particular for more deeply embedded planets. When the disk has a
strong toroidal magnetic field, horseshoe streamlines disappear and
the corotation torque is replaced by a so-called MHD torque that is
driven by angular momentum carried away by slow MHD waves excited in
the planet's vicinity. The recent work by \cite{Uribe15} shows that
this MHD torque can be positive and also reverse type I migration, in
agreement with earlier analytical and numerical works
\citep{Terquem03,Fromang05}.

{\bf (ii) Heating torque -- } Material accreted by a planet releases
energy that heats up the disk gas in the planet's vicinity (located
typically by a few Hill radii from the planet; the Hill radius is
defined at Eq.~\ref{eq:hill}). The gas flow near the planet implies
that this released energy forms two hot lobes: (i) one in front of the
planet in the azimuthal direction and inside the planet's orbit, and
(ii) a second one behind the planet in the azimuthal direction and
outside the planet's orbit \citep{Benitez15}. The gas density
decreases at the location of the lobes to maintain a mechanical
equilibrium with the surrounding gas. The under-dense lobe in front of
the planet favors inward migration (its gravitational force on the
planet has a negative azimuthal component), whereas the lobe behind
the planet favors outward migration. The flow asymmetry near the
planet, which is related again to the radial pressure gradient, makes
the lobe behind the planet hotter and more under-dense. The resulting
torque on the planet, termed "heating torque" by \cite{Benitez15},
therefore favors outward migration. It is not a corotation torque
since it is not associated with horseshoe U-turns. The magnitude of
the heating torque depends sensitively on the planet's growth
timescale and the disk opacity. The heating torque increases with the
accretion rate and with increasing disk opacity (since higher
opacities inhibit cooling). The results of \cite{Benitez15} show that
the heating torque can reverse type I migration of planets of 1 to 3
Earth masses that are located near 5 AU if their mass doubling time is
shorter than a few $\times 10^4$ years. Such short growth timescales
could be obtained through pebble accretion.

{\bf (iii) Eccentricity -- } Despite the relative efficiency of the
disk gas at damping eccentricities, a planet can maintain some level
of eccentricity if it is forced by another planet, in particular if
planets are in or near mean-motion resonances. \cite{Fendyke14} have
shown that the radial width of the planet's horseshoe region decreases
with increasing planet eccentricity, and therefore so does the
magnitude of the corotation torque. This clearly has important
implications for the strength of type I migration in multi-planetary
systems that contain Earth- to Neptune-mass planets.

{\bf (iv) Dynamical corotation torque -- } We have examined so far the
various contributions to the type I migration torque for planets on
fixed orbits, which assumes that the migration rate is directly given
by the disk torque, and that the latter is independent of the
migration rate. However, as a planet drifts in the disk, gas outside
the horseshoe region eventually executes a unique horseshoe U-turn
relative to the planet and thus contributes to the corotation
torque. The corotation torque imparted by orbit-crossing gas scales
with the mass flow rate across the orbit and therefore depends on the
migration rate. This torque is negative if the planet moves inward,
positive if it moves outward. It therefore causes a positive feedback
on migration.  It adds to the corotation torque due to the gas trapped
inside the planet's horseshoe region, which has two contributions.
First is the horseshoe drag caused by the gas in the horseshoe region
executing U-turns relative to the planet, as in the case when there is
no radial drift between the disk and the planet. Second is the torque
the horseshoe region has to apply on the planet so that both migrate
at the same rate.  This torque, which we call the trapping torque, is
perhaps more easily understood as the opposite of the torque the
planet exerts on its horseshoe region to have it migrate at its own
drift rate. The trapping torque is positive (negative) if the planet
moves inward (outward); it thus yields a negative feedback on
migration.

Adding all contributions together, we see that the corotation torque
has in general two components: (i) the {\it static corotation torque},
which is the horseshoe drag, and (ii) the {\it dynamical corotation
  torque}, which is the sum of the trapping torque and the
orbit-crossing torque \citep{MP03, SJP14}. As this terminology
implies, only the dynamical corotation torque depends on the migration
rate and causes a feedback on migration.

The key quantity to assess whether feedback is negative or positive is
a quantity called "vorticity-weighted coorbital mass deficit"
\citep{MP03}. This quantity is well known for massive planets that
open a gap around their orbit (see Sect.~\ref{ssec:mig_high}).  Since
low-mass planets do not or hardly deplete their horseshoe region,
above quantity should be thought of as a vorticity or vortensity
deficit, which comes about because orbit-crossing gas has a vortensity
different from the averaged vortensity of the horseshoe region
\citep{SJP14}. For type I migrating planets, the vortensity deficit
can be either negative or positive depending on the radial gradient of
disk vortensity in isothermal disk models (where the gas temperature
is kept stationary). Extension to non-isothermal, radiative disk
models (including viscous heating, stellar irradiation and radiative
cooling) has been carried out by \cite{Pierens15}, where the
vortensity deficit is primarily a surface density deficit brought
about by radial variation of the gas entropy in the disk; the
vortensity deficit can be positive or negative depending mostly on the
radial gradient of disk entropy. Conditions for type I migration to
runaway are described in \cite{SJP14} for isothermal disks, and in
\cite{Pierens15} for non-isothermal disks. For this review, it is
sufficient to take away that, in massive low-viscosity disks, the
dynamical corotation torque can dramatically slow down inward type I
migration (negative feedback), and possibly lead to runaway outward
type I migration (case where positive feedback causes a runaway). In
any case, the dynamical corotation torque can largely widen the range
of orbital radii where outward migration can occur
\citep{SJP14,Pierens15}.

%-------------------------------------
\subsection[]{Disk Migration of Massive Planets}
\label{ssec:mig_high}
%-------------------------------------
We provide in this section a short overview of planet-disk
interactions for massive planets that open a gap around their orbit,
their migration properties, and touch upon some aspects of their
observable signatures.

% -  -  -  -  -  -  -  -  -  -  -  -  -  -  -  
\subsubsection[]{Gap Opening}\label{ssec:gap_opening}
% -  -  -  -  -  -  -  -  -  -  -  -  -  -  -  
The more massive a planet, the stronger its wakes, and the closer to
the planet will wakes turn into shocks and deposit their energy and
angular momentum. As the inner wake carries negative fluxes of energy
and angular momentum, it will push gas inward, away from the planet's
orbit.  Similarly, the outer wake carries positive fluxes of energy
and angular momentum and will push gas outward, away from the
orbit. Massive planets thus progressively deplete their co-orbital
horseshoe region, which becomes an annular gap. The final gap width is
determined by the balance between the gap-opening gravity torque of
the planet and the gap-closing viscous and pressure torques
\citep{crida06}. In other words, the ability of a planet to carve a
gap does not only depend on the planet's mass, or, actually, the
planet-to-star mass ratio ($q=M_{\rm p} / M_{\star}$). It also depends
on the disk's aspect ratio, $h$, as $q/h^3$, and on the disk's
turbulent alpha viscosity, $\alpha$, as $\alpha h^2 / q$
\citep{LP86a,crida06}.

We stress that there is no accurate definition or criterion as to when
a planet opens a gap. The widely used gap-opening criterion formulated
by \cite{crida06} is for when the planet's co-orbital surface density
drops to 10\% its initial value. Of course this threshold value is
arbitrary, and planets less massive than predicted by Crida et al's
criterion can still open a partial gap or dip. This can be the case
for planets down to few Earth masses in weakly turbulent disks
\citep[e.g.,][]{rafikov02,Muto10, Dong11,Duffell13,Duffell15}. Also,
this criterion is for planets on fixed orbits. The gap-opening
criterion is modified by migration, particularly in massive disks, as
recently highlighted by \cite{Malik15}. An illustration of this is the
rapid inward migration expected for planets formed by disk
fragmentation \citep[e.g.,][]{BMP11}, which is initially so rapid that
planets have no time to carve a gap, unless or until significant gas
accretion occurs \citep{Zhu12,NayakshinCha13,Stamatellos15}. We also
point out the recent work by \cite{Fung14} who estimated via
hydrodynamical simulations the gas surface density contrast inside and
outside a planet gap. Note that simulations of MHD turbulent disks
show that the width and depth of planet gaps can be somewhat different
from viscous disk models \citep{Papa04, Zhu_inv}.

Gap-opening planets are often invoked to explain the large cavities
observed in transition disks \citep[see, e.g., the review
by][]{Espaillat14}.  It should be borne in mind that the width of a
planetary gap varies with dust size. Dust grains up to few tens of
microns are tied to the gas, so the width of their gap is just as
narrow as in the gas. Dust grains in the mm/cm range (depending on the
local gas density) are generally not tied to the gas: they undergo
significant radial drift and tend to concentrate in regions of
pressure maxima (see Sect.~\ref{ssec:pebaero}). Simulations of
planet-disk interactions show that the outer edge of a planet gap is
generally a robust pressure maximum but the inner edge is not. This
implies that mm/cm-grains initially beyond the planet's orbit can be
trapped efficiently at the gap's outer edge, and those initially
inside the planet's orbit will migrate inward \citep[e.g.,][]{Zhu12}.
A single massive planet could therefore simultaneously open a narrow
gap in $\mu$m-grains and a large cavity in mm-grains. We point out
that large cavities in $\mu$m-grains are not a natural expectation of
planet-disk interactions, unless one invokes the presence of several
massive planets in the disk \citep{Zhu11}. Finally, we stress how
uncertain it would be to apply predictions of gap widths and depths
for the gas to observed gap structures in protoplanetary disks imaged
in the mm/submm, like for example the (suggestive) gap structures seen
in the HL Tau disk imaged by ALMA \citep{HLTau}.

Quite similarly, massive, yet unseen planets have been widely invoked
to explain the spiral density waves observed in near-infrared images
of protoplanetary disks. While spiral waves could be generated by
other means, for example by gravitational instability, studies on the
observability of the spiral wakes induced by a massive planet,
coupling hydrodynamical simulations and radiative transfer
calculations, are emerging \citep{Juhasz15,Zhu15spirals}.  The work by
\cite{Dong15spirals} provides convincing evidence that the two spirals
seen in MWC 758 in polarized scattered light \citep{Benisty15} can be
due indeed to an unseen outer planet companion of a few Jupiter
masses.

% -  -  -  -  -  -  -  -  -  -  -  -  -  -  -  
\subsubsection[]{Formation of a Circumplanetary Disk}\label{ssec:circum}
% -  -  -  -  -  -  -  -  -  -  -  -  -  -  -  
Part of the gas that is progressively expelled from the planet's
co-orbital region during the opening of a gap forms a circumplanetary
disk. The recent discoveries of young, massive planets on wide orbits
by direct imaging, like HD 100546 b (recently confirmed by
\citealp{Currie14} and \citealp{Quanz15}), has stimulated a number of recent works
that have examined the structure, accretion rate and observability of
circumplanetary disks. Accretion onto the protoplanetary disk is found
to be inherently 3D and to proceed from high latitudes
\citep[e.g.,][]{Tanigawa12,Szulagyi14}, can be stochastic and even
launch outflows \citep{Gressel13}. The observability of
circumplanetary disks has been assessed by the prediction of distinct
features in the gas kinematics \citep{Perez15} and spectral energy
distribution \citep{Zhu15CPD,Eisner15,Montesinos15}.

% -  -  -  -  -  -  -  -  -  -  -  -  -  -  -  
\subsubsection[]{Type II Migration}\label{ssec:type2}
% -  -  -  -  -  -  -  -  -  -  -  -  -  -  -  
The migration of massive, gap-opening planets is driven by the wake
torque and the corotation torque (both its static and dynamical
components).  The balance between both torques depends on how deep and
wide the gap is, which, as we have seen above, depends on the planet's
mass, the disk's aspect ratio and viscosity. For planets that open a
deep gap around their orbit, which, roughly speaking, correspond to
planets more massive than Jupiter, the corotation torque is largely
suppressed and the wake torque is weakened. Still, the wake torque
remains the main driver of migration. This corresponds to the
so-called type II migration regime. It is directed inwards and runs on
timescales longer than $10^{4-5}$ yrs.  Recently, \cite{DK15} have
shown through numerical simulations that type II migration does not
depend on the viscous inflow speed of the disk gas. This implies that
type II migration is not tied to the disk's viscous evolution, as
often thought, and that gas can cross the gap during the planet's
migration. Type II migration is considerably slowed down when the
planet mass becomes larger than the mass of the gas outside the planet
gap.

% -  -  -  -  -  -  -  -  -  -  -  -  -  -  -  
\subsubsection[]{Type III Migration}\label{ssec:type3}
% -  -  -  -  -  -  -  -  -  -  -  -  -  -  - 
For planets that open a partial gap around their orbit, which is
typically the case of Saturn-mass planets, the corotation torque still
plays a major role in the migration. At low to moderate gas surface
densities at the planet's location, that is when the local Toomre
parameter $Q \equiv c_{\rm s}\Omega_{\rm K} / \pi G \Sigma$ typically
exceeds $10$, the migration regime is intermediate between the type I
and type II regimes \citep[e.g.,][]{MP03,CM07,DAngelo08}. At larger
gas surface densities at the planet's location ($Q \lesssim 10$), the
dynamical corotation torque becomes important and leads to a different
migration regime termed type III migration \citep{MP03}. The
co-orbital vortensity deficit featured by the dynamical corotation
torque (see Sect.~\ref{ssec:mig_progress}) becomes a surface density
deficit or mass deficit (it is the mass that should be added to the
planet's horseshoe region so that it gets the averaged surface density
of the orbit-crossing gas). When the mass deficit becomes larger than
the mass of the planet plus its circumplanetary disk, migration runs
away, otherwise the dynamical corotation torque accelerates migration
but does not cause a runaway \citep{MP03}.  The planet's
circumplanetary disk can have a significant impact on migration
\citep{gda2005,Peplinski1,CBKM09}.

%-------------------------------------
\subsection[]{Disk Migration and the Observed Exoplanets}
\label{ssec:mig_obs}
%-------------------------------------
This section discusses the role of planet-disk interactions in some of
the observed properties of planetary systems. Sect.~\ref{ssec:obliq}
deals with spin-orbit (mis-)alignments among hot Jupiters and their
constraints on planet migration scenarios. Sect.~\ref{ssec:multi}
discusses the relevance of planet-disk interactions for the near- and
non-resonant architecture of many of the multi-planet systems
discovered by {\it Kepler}. For a detailed statistical comparison
between the orbital properties of exoplanets and those predicted by
global models of planetary formation and evolution, the reader is
referred to, e.g., \cite{BenzPP6} and \cite{Mordasini15review}.

% FFFFFFFFFFFFF
\begin{figure}
\centering
\includegraphics[width=\hsize]{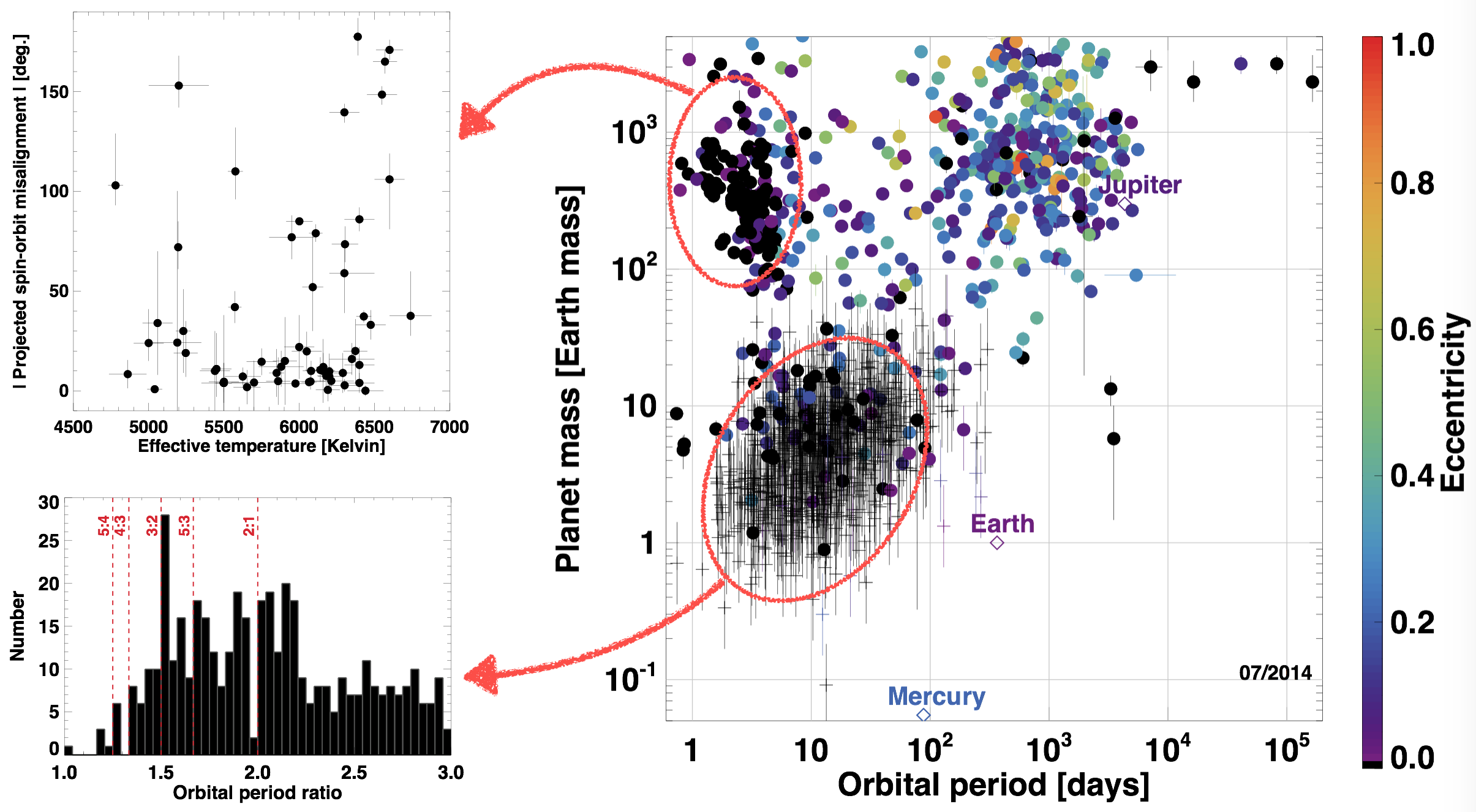}
\caption{Right: Mass, orbital period and eccentricity of observed
  exoplanets (filled circles), and for Mercury, Earth and
  Jupiter. Planets labelled by plus signs are planets detected by
  transit only, for which the mass is inferred from the mass-radius
  relationship $M \propto R^{2.06}$ for low-mass planets with both
  mass and radius determined. A 50\% error bar on the mass has been
  added to these planets detected by transit only. Upper-left:
  projected obliquity versus stellar effective
  temperature. Lower-left: period ratio histogram for confirmed
  multi-planet systems, the period ratio of a few mean-motion
  resonances is shown by vertical dashed lines. All data were
  extracted from \href{http://exoplanets.org}{exoplanets.org}.}
\label{fig:CB3}
\end{figure}
% FFFFFFFFFFFFF

% -  -  -  -  -  -  -  -  -  -  -  -  -  -  -  
\subsubsection[]{Aligned and Misaligned Hot Jupiters: Constraints on Planet Migration Scenarios}
\label{ssec:obliq}
% -  -  -  -  -  -  -  -  -  -  -  -  -  -  -  
Observations of sky-projected planet obliquities have gained
considerable importance in the past few years (the obliquity of a
planet is the angle between the star's spin axis and the planet's
orbital axis). Projected obliquities have been primarily obtained via
the Rossiter MacLaughlin effect \citep[e.g.,][]{Winn05}, which is a
time-varying distortion in the spectral lines of the star caused by a
transiting planet. Other techniques, like the analysis of starspots
occultations by a planet, have also been used
\citep[e.g.,][]{SanchisOjeda11}. Projected obliquities have been
determined for a little more than 60 planets at the time of writing,
the vast of majority of them are hot Jupiters with orbital periods
below 10 days.  Surprisingly, a fair number of hot Jupiters have a
large projected obliquity (see Figure~\ref{fig:CB3}). When their
projected obliquity exceeds 90\degree, planets are said to be
retrograde. It is seen that the projected obliquities seem to depend
on the star's effective temperature, with hot Jupiters around hot
stars being more likely misaligned \citep[see, e.g., the review
by][]{WinnARAA15}.  These observations have considerable impact on
evolutionary models of planetary systems, since they indicate that
dynamical processes may commonly lead planets to acquiring large
obliquities, which is often thought of as planets getting large
inclinations at some point of their evolution. We mention that, under
some circumstances, measuring the stellar inclination relative to the
line of sight, $i_{\star}$, for example through asteroseismology, can
constrain the true obliquity of a planet (a small $i_{\star}$ implying
a large misalignment; \citealp{Huber13,Chaplin13}) and even determine
it \citep{Benomar14}.  At the time of writing, true obliquities have
been estimated for 11 planets\footnote{See, e.g.,
  \href{http://www.astro.keele.ac.uk/jkt/tepcat/rossiter.html}{http://www.astro.keele.ac.uk/jkt/tepcat/rossiter.html}.},
they exceed 30\degree~for 5 of them.

How to explain these observations? Hot Jupiters are thought to form at
typically several AU from their central star, reaching orbital periods
of a few days either by disk migration or by high-eccentricity
migration. Disk migration is expected to preserve zero obliquity
\citep[e.g.,][]{Bitsch13b} unless (i) the disk becomes misaligned with
the stellar equator due to gravitational interactions with nearby
stars \citep{BLP10,Batygin12,Picogna15}, or (ii) the star itself
becomes misaligned, which might occur through star-planet tidal
interactions (tidal flip of stellar axis, \citealp{Cebron13,
  Barker14}). Note that the delivery of hot Jupiters by disk migration
does not have to proceed solely by smooth type II migration.  It may
be assisted by the inward scattering of a giant planet resulting from
interactions with one or more giant planet companions in the
protoplanetary disk \citep[e.g.,][]{Marzari10}.  In the
high-eccentricity migration scenario, a Jupiter-mass planet begins on
an inclined, highly eccentric orbit that shrinks and circularizes due
to tidal interactions with the central star. Mechanisms that may pump
planet eccentricities and/or inclinations in the first place include
(i) planet-planet scattering between two or more massive planets, a
scenario that has been proposed to explain the broad distribution of
exoplanet eccentricities
\citep[e.g.,][]{RasioFord96,Chatterjee08,Juric08}, and (ii) Kozai
cycles with an inclined, probably stellar-mass companion
\citep[e.g.,][]{WuMurray03,FabryckyTremaine07,Naoz12}.
High-eccentricity migration can produce a broad range of planet
obliquities depending on the excitation mechanism and the
effectiveness of tidal interactions at damping obliquities (which
depends on the star's mass, rotation period, the planet's mass and
orbital separation etc.).

Interestingly, it has been suggested that perhaps {\it all} hot
Jupiters formed by high eccentricity migration with a broad range of
initial obliquities, and that tidal interactions with the central stars
were more or less efficient at reducing obliquities over the age of
the systems (that is, at aligning the stellar spin axis with the
planet's orbital axis, \citealp{Triaud10, Albrecht12}).
\cite{Albrecht12} argued that aligned systems would be those where the
timescale for spin-orbit alignment is shorter than the age of the
system. They used the estimates of \cite{Zahn77} for the tidal
synchronization timescales in close binary stars, showing that tidal
alignment should be more efficient for cooler stars that harbor a
convective envelope. While Zahn's theory does predict an increase in
the alignment timescale with the star's effective temperature, in
practice it gives alignment timescales that are at least one or two
orders of magnitude longer than the age of the systems where projected
obliquities have been measured \citep{Albrecht12,Ogilvie14}. This
shows the limitations in applying Zahn's theory of tidal evolution of
binary stars to hot Jupiters. This result has nevertheless stimulated
much work on star-planet tidal interactions. In particular, the
conditions for tides to produce alignment faster than orbital decay
lead to the prediction that there should be as many retrograde aligned
planets as prograde aligned ones, which is not what is observed
\citep{Lai12, RogersLin13}.

Could measurements of projected obliquities help distinguish and/or
constrain the mechanisms of hot Jupiter migration? Answer to this
important question probably requires a statistical comparison between
model predictions and observations. We mention the recent work by
\cite{CridaBatygin14}, who compiled the predictions of
high-eccentricity migration models in terms of projected obliquities,
and compared them with observations. They find that the models
successfully reproduce the distribution of projected obliquities
beyond 40\degree, but they all underestimate the proportion of systems
with low obliquities. An alternative scenario mentioned above is the
possibility that the disk can be misaligned relative to the star's
equator due to the interaction with a binary companion
star. \cite{CridaBatygin14} use Monte-Carlo simulations to predict the
distribution of projected obliquities according to this scenario. They
find that, (i) just like for the combined high-eccentricity and tidal
friction scenario, the misaligned disk migration scenario can robustly
reproduce the distribution of projected obliquities beyond 40\degree,
but (ii) only the aligned disk migration scenario can account for the
number of aligned and nearly aligned hot Jupiters, depending on the
semi-major axis distribution of the binary companions. Inclusion of
magnetic interactions between the disk and the central star allows the
misaligned disk migration scenario to reproduce the observed trend
between misalignment and stellar mass \citep{SpaldingBarygin15}. These
comparisons highlight that (i) current observations cannot disentangle
between the high-eccentricity migration scenario and the disk
migration scenario in a misaligned disk to explain misaligned hot
Jupiters, and that (ii) delivery of aligned hot Jupiters via disk
migration is necessary to account for the number of aligned hot
Jupiters. Chemical depletions (e.g., O and C abundances) in the
atmosphere of hot Jupiters could be a way forward in constraining
their migration scenario \citep{Madhu14}. We also point out that the
fraction of observed aligned hot Jupiters does not seem to vary
significantly with orbital period\footnote{This is inferred from the
  sample of planets listed in
  \href{http://exoplanets.org}{exoplanets.org} that have a minimum
  mass greater than 0.5 $M_{\rm J}$. Defining as aligned a planet with
  projected obliquity less than about 40\degree, we find that, in the
  range of orbital periods between 1 and 3 days, there are 18 aligned
  and 7 misaligned planets, while for orbital periods between 3 and 5
  days, there are 16 aligned and 8 misaligned planets.} which, in our
opinion, is a challenge to the above tidal alignment theory.  We
finally mention the recent work by \cite{Mazeh15}, who show that the
trend of lower obliquity around cooler stars extends to {\it Kepler}
objects of Interest with orbital periods up to at least 50 days, for
which star-planet tidal re-alignment should be largely negligible.

% -  -  -  -  -  -  -  -  -  -  -  -  -  -  -  
\subsubsection[]{Architecture of {\it Kepler}'s Multi-Planetary Systems}
\label{ssec:multi}
% -  -  -  -  -  -  -  -  -  -  -  -  -  -  -  
{\it Kepler} has uncovered a large population of Earth- to
Neptune-mass planets in compact multi-planetary systems with orbital
periods between 1 and 100 days. The distribution of period ratios
between all pairs of planets of a given system is displayed in the
lower-left corner of Figure~\ref{fig:CB3} for {\it Kepler}'s confirmed
multi-planet systems. There are two salient points. First, there are
many planet pairs far from resonances, while the usual expectation
from disk migration models is that {\it Kepler}-mass planets should
form resonant planet pairs, that is, one would a priori expect only
spikes in this distribution at the location of the resonant period
ratios, like 2:1, 3:2 etc. Second, planet pairs near resonances tend
to have period ratios somewhat greater than resonant. It is worth
noting that, despite a much smaller number statistics, a similar trend
can be seen for planetary systems detected by RV techniques only. We
point out that some features in the period ratio distribution may be
artifacts of unseen additional companions \citep{Steffen13miss}. For
example, a two-planet system with a period ratio $\sim1.9$ could
actually be a three-planet resonant system where the middle planet,
missed by the transit observation, is in 4:5 mean-motion resonance
with the innermost planet, and in 3:2 mean-motion resonance with the
outermost planet.

How to explain these observed features? One model that has been
recently put forward is in-situ growth. Here, it is important to
stress that this is not in-situ growth all the way from
micrometer-sized grains to planetary-sized objects, but in-situ growth
of pre-formed and pre-evolved Mars- to Earth-sized embryos in a
gas-free disk, and that these embryos may well have been delivered
previously by disk migration, before the protoplanetary disk gets
cleared out. The period ratio distribution predicted by the in-situ
growth model of \cite{Hansen13} tends to produce too few close and
resonant planet pairs compared to observations. The recent work by
\cite{ogiharamorbidelli2015} shows that planet embryos actually form
very rapidly in the inner parts of protoplanetary disks, and that
therefore disk migration cannot be neglected. Inclusion of disk
migration in the "in-situ" growth calculations of
\cite{ogiharamorbidelli2015} shows that disk migration leads this time
to an excess of close pairs of planets compared to observations.

Another model that has been highlighted is that a resonant planet pair
formed by disk migration may evolve away from resonance due to
star-planet tidal interactions, which tend to slowly increase the
planets' period ratio over time, at a rate that depends on the
star-planet separation \citep{Papa11, LithwickWu12}. While tides may
indeed account for some planet pairs being just wide of resonance, we
note that a similar trend is observed for planet pairs above 10 days,
for which tides should be inefficient \citep{BP13}.

More recently, disk migration of planet pairs has been
revisited. \cite{BP13} have shown that, at the typical location of
Kepler planets, the wakes of the planets should be shock waves, which
implies that: (i) the planets should open partial gaps around their
orbits, and (ii) the wakes deposit energy and angular momentum in the
co-orbital region of each planet, with the consequence that the
planets tend to repel each other over time. This stresses that disk
migration models do not necessarily predict {\it Kepler}-mass planets
to form resonant pairs.  Turbulence in the disk is another way to
explain why many low-mass multiple planet systems are near- or
non-resonant \citep{PBH11,Rein12b}, and may play a prominent role in
the formation of closely packed systems \citep{Paardekooper13}.

%==============
\section{Internal Evolution of Planets}\label{sec:theevolutionofplanets}
%==============
We now turn our attention to individual planets, treating them no more
just as point-like masses, but as geophysical objects that are
characterized by a radius, a luminosity, an atmospheric and a bulk
compositions, which can all evolve in time.

A small fraction of self-luminous objects like $\beta$ Pic b
\citep{lagrangegratadour2009} among the known extrasolar planets has
been directly detected at young ages, using coronagraphs and dedicated
data reduction procedures, and a handful of objects has even been
observed during their formation, which occurs on a timescale of
$\sim10^{7}$ years \citep[e.g.,][]{Quanz15}.  However, the large
majority of known exoplanets orbits stars that are at least $\sim$100
Myrs old with a typical age of several Gyrs, meaning that we usually
observe (exo)planets at an epoch much after their formation. Many of
these objects are then too faint or too close to their star to be
detected directly. More often their mass can, however, be inferred
from radial velocity measurements.  Further, if the planets are
transiting their star, it becomes possible to measure their radius if
the radius of the host star is known. Transits also allow to probe the
planetary atmospheres and to get hints at the planet's abundances and
day-side emission. The combination of radial velocity and transit
measurements finally yields the mean density of a planet, which is its
first rough but important geophysical characterization
\citep[e.g.,][]{dressingcharbonneau2015,santosadibekyan2015}.

The number of planets for which various observational data is
available has increased enormously in the past two decades
\citep[e.g.,][]{mayormarmier2011,boruckikoch2011,marcyisaacson2014}. One
of the major goals of the research on (exo)planets is to better
understand the physics of planet formation thanks to this data,
including in particular the Solar system. A typical approach is to
compare the results from theoretical planet formation models and the
observational constraints from the aforementioned techniques.
However, in order to draw conclusions about the formation of planets
from properties that we observe today like orbital parameters, mass,
radius, luminosity and composition, one needs to understand not only
how planets form, but also how they evolve from formation until the
epoch of observation. The reason for this is that the evolution spans
Gyr-long timescales, and the link between observation and formation
can be non-trivial and potentially mask the original imprints of
formation.

Close-in low-mass planets are an illustrative example. Their formation
mechanism is currently debated (see, e.g.,
Sect.~\ref{ssec:multi}). Additional constraints that can be derived
from their geophysical characteristics like for instance the H/He
envelope mass are therefore very valuable to distinguish different
formation scenarios (Sect.~\ref{sect:enveevap_mrho_sol_gas}).
However, because of envelope evaporation (atmospheric escape), some
planets might have started with a gaseous envelope after formation but
then lost it during the evolutionary phase.

Directly imaged planets are another example. Even for these young
planets, which are typically only a few Myrs old, the question of
understanding their temporal evolution since formation is of central
importance when it comes to infer their mass based on their
luminosity. This mass-luminosity relationship is in turn of high
interest for formation theories as will be discussed in
Sect.~\ref{sect:ini_cond}: it constrains the physics of gas accretion
and potentially even the mode of formation, namely core accretion
\citep{perricameron1974,mizuno1980,bodenheimerpollack1986} or
gravitational instability \citep{toomre1981,Boss97,durisenPP5}.  Not
only must evolutionary models be well-suited to describe the governing
physics, but also the correct initial conditions for the planet's
evolution must be known, meaning that formation and evolution must be
linked self-consistently.

In this section, we mostly concentrate on how to derive constraints on
formation models from studying the long-term internal (mainly
thermodynamical) evolution of planets after the dissipation of the
protoplanetary disk, which is the link between formation and
observation. Concentrating mainly on planets with primordial H/He, we
address some of the most important aspects of planetary evolution like
cooling and contraction, the mass-luminosity relation, and the bulk
composition as expressed in the mass-radius and mass-mean density
relations. We also quickly touch upon other topics that can play
important roles like atmospheric escape, radius inflation and
deuterium burning. For further aspects of the theory of planetary
evolution, including crucial elements like high-pressure physics and
equations of state, the internal rotation of planets and their shape,
the specific properties and cooling histories of the Solar system
planets, or potential phase separations of elements in planetary
interiors (to name just a few), the reader is referred to the reviews
of
\citet{stevenson1982b,guillot2005,fortneynettelmann2010,chabrierjohansen2014,
  guillotgautier2014} or \citet{baraffechabrier2014}.

%-------------------------------------
\subsection{Thermodynamical Evolution: Cooling and Contraction}
% -------------------------------------
The thermodynamical evolution of giant planets after formation at
constant mass is controlled by cooling and contraction. They evolve
from a comparatively hot post-formation state characterized by high
entropy, high luminosity, low degeneracy, and large radius to a colder
state with low entropy, low luminosity, higher degeneracy, and smaller
radius \citep[e.g.,][]{guillot2005}.  In contrast to stars, due to
their comparatively low internal temperature, giant planets do not
radiate away energy released by thermonuclear fusion as they are not
massive enough to allow for fusion in their interior. The only
exception may be a short stage of deuterium burning in planets with
masses in excess of 12-13 $M_{\rm J}$, as we will see in
Sect.~\ref{subsect:dburn}. But, for the bulk of giant planets, cooling
and contraction are the main source of internal luminosity
\citep{hubbard1980}, which can be seen from the fact that Jupiter,
Saturn and Neptune emit more energy than they receive from the Sun.

The total amount of energy that can be radiated away by a giant planet
during its lifetime is primarily determined by its post-formation
entropy (Sect.~\ref{sect:ini_cond}). For terrestrial planets, such as
Earth, another important contributor to the internal luminosity,
besides delayed secular cooling, is the radioactive decay of unstable
heavy element isotopes \citep[e.g.,][]{urey1955,hofmeister2005}. For
the Earth, the total intrinsic heat flux is around 47 TW (which is
only about 0.03\% the solar irradiation flux), of which about half is
estimated to arise from radiogenic heating \citep{gandogando2011}. For
Jupiter, the intrinsic luminosity (which is of the same order as the
solar irradiation flux, \citealp{guillotgautier2014}) is largely
dominated by the cooling and contraction of the fluid H/He envelope; a
possible solid core could contribute a few percent, while the
radiogenic contribution is tiny ($\sim 10^{-4}$).

The cooling of gas giant planets is caused by the radiation escaping
through their atmosphere, which decreases the total energy of the
planet. Because of virial equilibrium, for self-gravitating objects
composed mainly of ideal gas, radiative escape causes somewhat
counter-intuitively the interior to heat up
\citep[e.g.,][]{guillot2005}.  Such objects are therefore said to have
a negative heat capacity: the loss of energy at their surface leads to
an increase in their internal energy and central temperature, which is
compensated for by the object's contraction. This regime applies to
low-mass pre-main-sequence stars, as well as extremely young hot giant
planets at ages less than $10^{5-6}$ years they are not yet
significantly degenerate \citep{graboskeolness1975}.

The situation is quite different for older giant planets however, for
which cooling and contraction cause the interior temperature to
decrease. This is because the thermal pressure needed to balance the
planet's gravity is established by degenerate electrons, which
increase their mean kinetic energy, in contrast to ions, which
decrease their kinetic energy and cool, yielding most of the intrinsic
luminosity \citep{guillot2005}.  In this phase, giant planets behave
qualitatively like degenerate brown dwarfs \citep{graboskeolness1975}.

The time evolution of the cooling of a planet, and thus its internal
evolution, depend on how energy is transported from the interior to
the top and then through the atmosphere of the planet.  The atmosphere
can be thought of as a blanket which, with increasing thickness (i.e.,
optical depth), diminishes the planet's ability to cool, making it
stay hot and have a larger physical radius for a longer period of time
at high opacity \citep{burrows2007}.  For strongly irradiated planets
the cooling timescale also increases: the incoming radiation makes the
atmospheres more isothermal, reducing the effectiveness of radiative
energy transport, as the atmospheric temperature gradient decreases
\citep{burrows2000,guillot2002}.

% FFFFFFFFFFFFF
\begin{figure}[t]
\centering
\begin{minipage}{0.48\textwidth}
\includegraphics[width=1.0\textwidth]{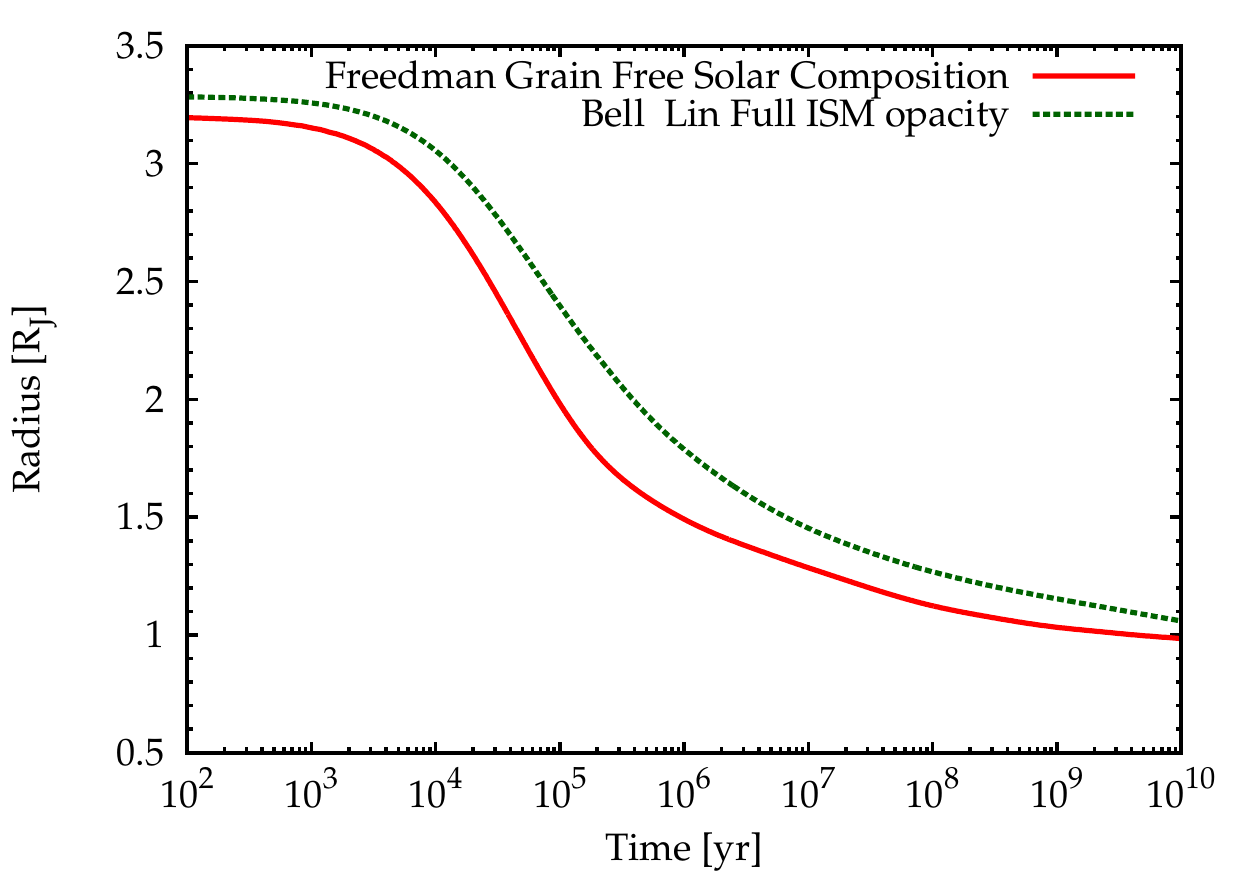}
\end{minipage}
\begin{minipage}{0.03\textwidth}
\end{minipage}
\begin{minipage}{0.48\textwidth}
\includegraphics[width=1.0\textwidth]{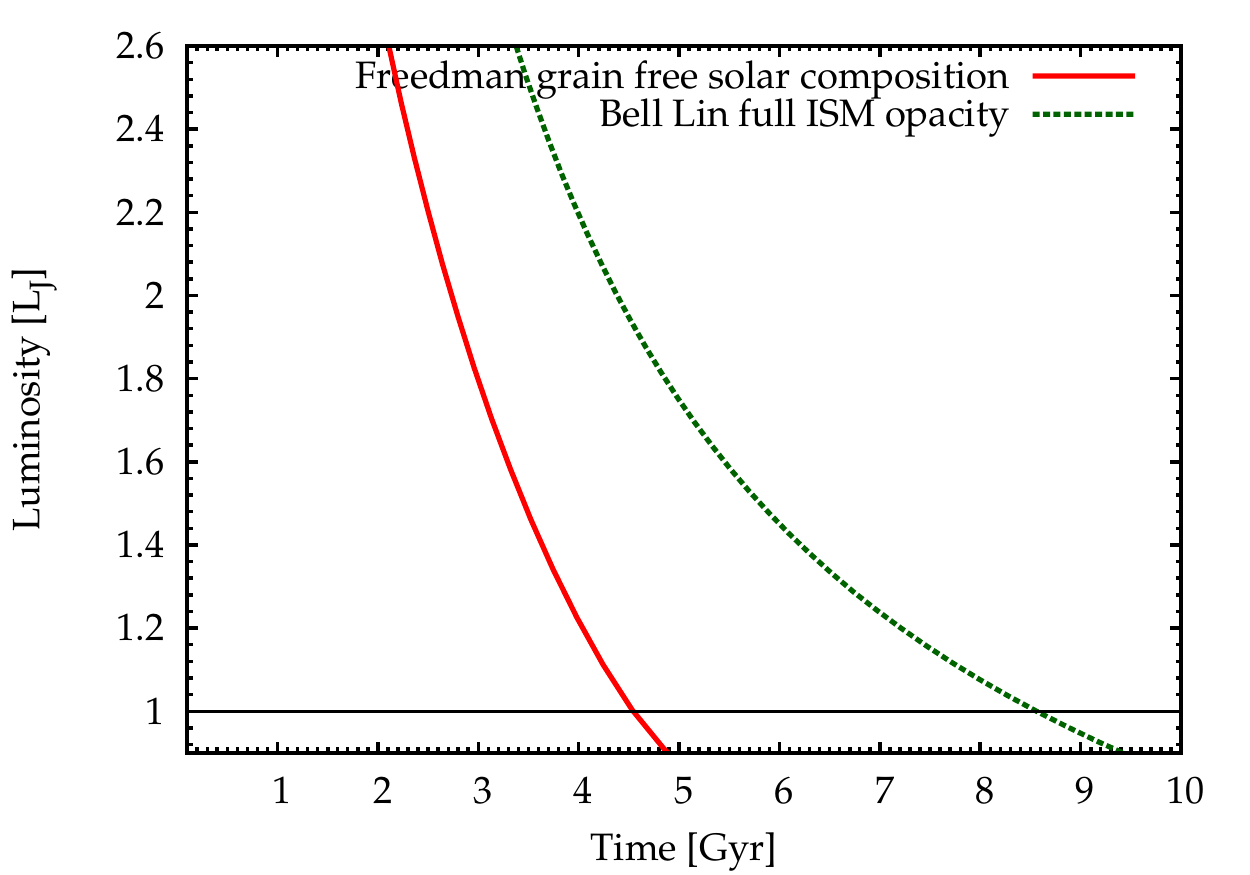}
\end{minipage}
\begin{minipage}{0.48\textwidth}
\includegraphics[width=1.0\textwidth]{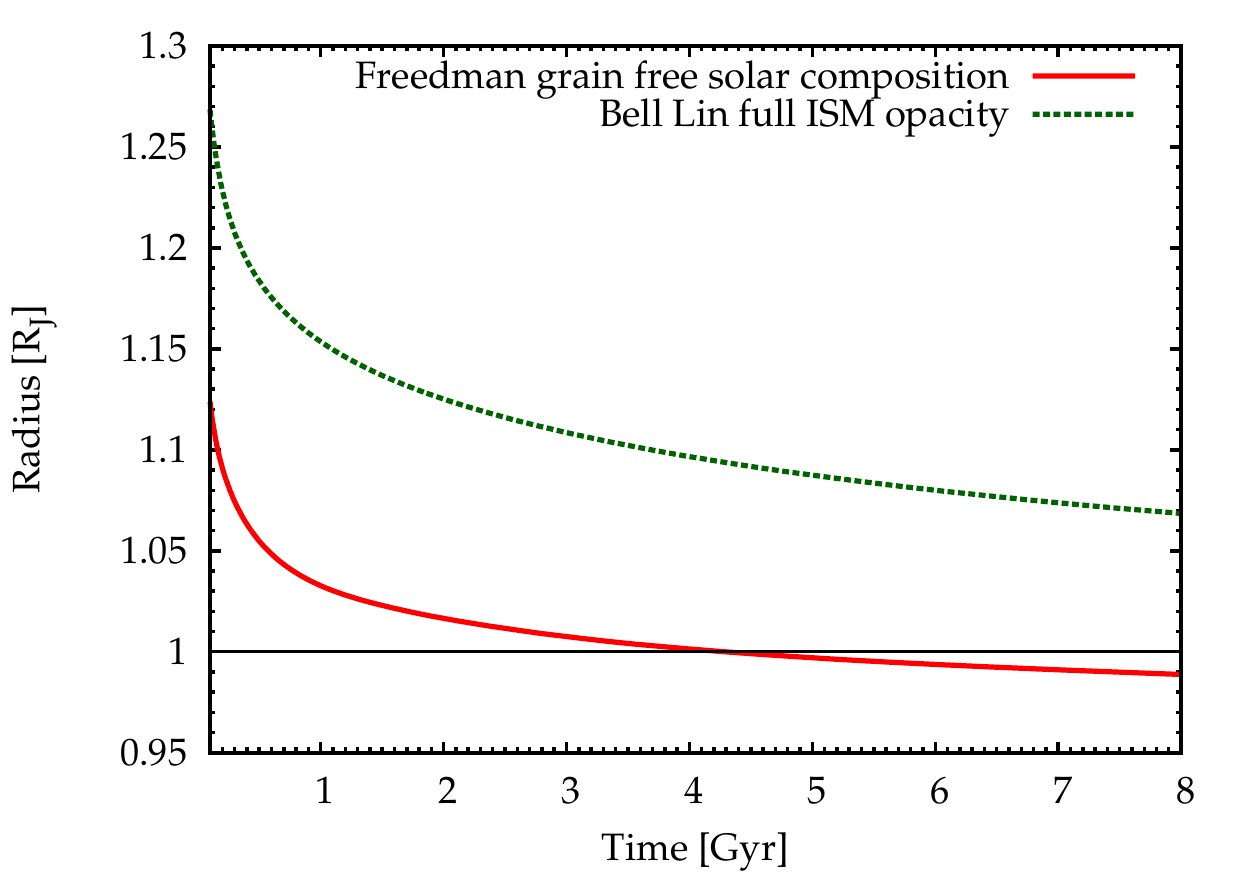}
\end{minipage}
\begin{minipage}{0.03\textwidth}
\end{minipage}
\begin{minipage}{0.48\textwidth}
\includegraphics[width=1.0\textwidth]{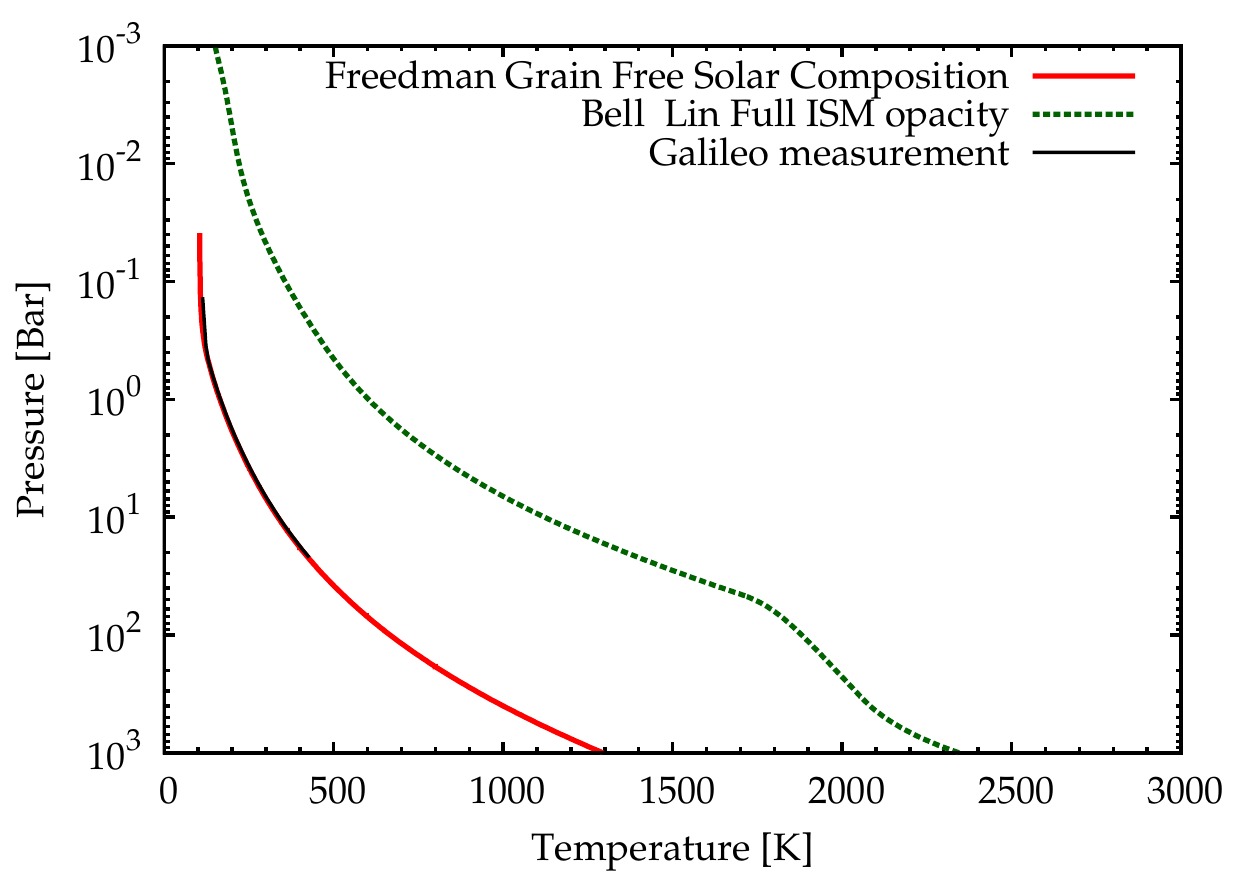}
\end{minipage}
\caption{Calculation of the internal evolution of a Jupiter-mass
  planet for two different input opacities: full ISM grain opacities
  \citep[][green dotted curves]{bell1994}, and Solar abundance
  molecular opacities \citep[][red solid curves]{freedman2008}. The
  time evolution of the planet's $\tau=2/3$ radius is shown in the
  upper- and lower-left panels with two different time windows. The
  time evolution of the planet's luminosity is in the upper-right
  panel, and the pressure-temperature profiles of the planet's
  atmosphere at Jupiter's current age is displayed in the lower-right
  panel. The black solid curve in the lower-right panel shows the
  pressure-temperature profile measured by the Galileo probe
  \citep{seiff1996}.}
\label{fig:opa_evo}
\end{figure}
% FFFFFFFFFFFFF
An example of planetary evolution calculation illustrating the impact
of opacities is shown in Figure~\ref{fig:opa_evo}
\citep{mordasinialibert2012b}. For such calculations, the fundamental
equations for a planet's internal structure (equations of mass
conservation, hydrostatic equilibrium, energy conservation, and energy
transport by radiation or convection) are solved
\citep[e.g.,][]{guillotgautier2014} which allows the construction of
evolutionary sequences based on the principle that the luminosity is
equal to minus the temporal change of the planet's total energy. The
post-formation evolution of a Jupiter-like planet is simulated
assuming (ad hoc) either very high opacities caused by ISM grains, or
more realistic grain-free molecular opacities for a solar composition
gas which are about three orders of magnitude lower (see
\citealp{bell1994} and \citealp{freedman2008} for the grain and
molecular opacities, respectively).  For both sets of opacities, the
planet has the same initial entropy. Note that planets with the same
set of opacities, but different initial entropies, should still get
the same structure in the end because the initial evolutionary
timescales are short, at least if the post-formation entropies are not
very low (Sect.~\ref{sect:ini_cond}). Figure~\ref{fig:opa_evo}
displays the time evolution of the planet's radius and luminosity, as
well as the pressure-temperature profiles of the planet's atmosphere
at Jupiter's current age. One clearly sees that for ISM grain
opacities (in reality the grains would fall out of the atmosphere on a
very short timescale, \citealt{mordasini2014}), the planet's radius
remains larger than for molecular opacities. Also, after a few Gyrs,
the planet luminosity obtained with ISM grain opacities is about a
factor 2 larger than with molecular opacities. This is expected, as
with ISM opacities the planet cools much more slowly, so that at late
evolutionary times the planet has retained more heat than if it had
cooled down with low molecular opacities.  Only molecular opacities
can reproduce Jupiter's present day radius, luminosity, and the
pressure-temperature profiles measured by the Galileo probe
\citep{seiff1996}.

Energy transport in the interior of gaseous planets is commonly
assumed to be due to convection. This is because the interior is
usually characterized by high opacities, such that Schwarzschild's
criterion for convection is satisfied \citep[but see
also][]{guillotchabrier1995}. However, different means of energy
transport might make planetary interiors considerably deviate from the
picture of efficient convective energy transport. If planetary
envelopes have a compositional gradient with mean molecular weight
increasing towards the interior, the large-scale energy transport by
convection could be shut down. The reason for this is that blobs of
gas perturbed to upper layers of the atmosphere must sink back
downward because the surrounding material has a lower density due to a
smaller mean molecular weight. The possible implications for planetary
structures were discussed by \citet{stevenson1985} and recently
revisited by \citet{leconte2012}.  Following these studies, layers
which are unstable according to Schwarzschild's criterion, but stable
when including the compositional gradient, are thought to become
semi-convective, exhibiting many convective layers separated by thin
conductive layers with steep compositional gradients. The conductive
energy transport is established by the thermal motion of electrons and
ions.  Planets with such semi-convective envelopes are thought to cool
much slower, as the conductive boundary layers hamper efficient energy
transport. It has been suggested that this phenomenon could be
responsible for Saturn's high luminosity \citep{leconte2013}, which is
too high when compared to classical, fully convective models
\citep{pollack1977}. Semi-convection is thus an alternative to the
classical explanation that Saturn's excess luminosity is caused by the
phase separation of hydrogen and helium (He insolubility) and a
subsequent helium ``rain'' into the deeper layers, which releases
gravitational potential energy \citep[e.g.,][]{stevensonsalpeter1977}.

There are two important processes to assess the extent to which strong
compositional gradients, and therefore semi-convection, are present in
planetary envelopes. First, the solid core present in planets thought
to have formed by core accretion might dissolve, leading to an
increasing mean molecular weight in the layers close to the
core-envelope boundary.  This has been studied by \citet{guillot2004}
and \citet{wilson2012}, who found that a dissolution and subsequent
mixing of the core might be possible. Second, during the planet
formation phase, the destruction and heavy element deposition by
planetesimals breaking up in the envelope similar to comet
Shoemaker-Levy 9 could also lead to compositional variations. The
ablated material could either stay suspended near the location where
it has been deposited, settle towards deeper regions, or get mixed
through the envelope by convection \citep{mordasini2006,iaroslaw2007}.
Clearly, this can lead to compositional gradients in the planet's
envelope that are determined by the formation process. In this way the
planetesimal accretion history might influence the planet's evolution
even at late times.

%-------------------------------------
\subsection{Initial Conditions for Planetary Evolution}
\label{sect:ini_cond}
\begin{figure}[t]
\centering
\includegraphics[width=1.\textwidth]{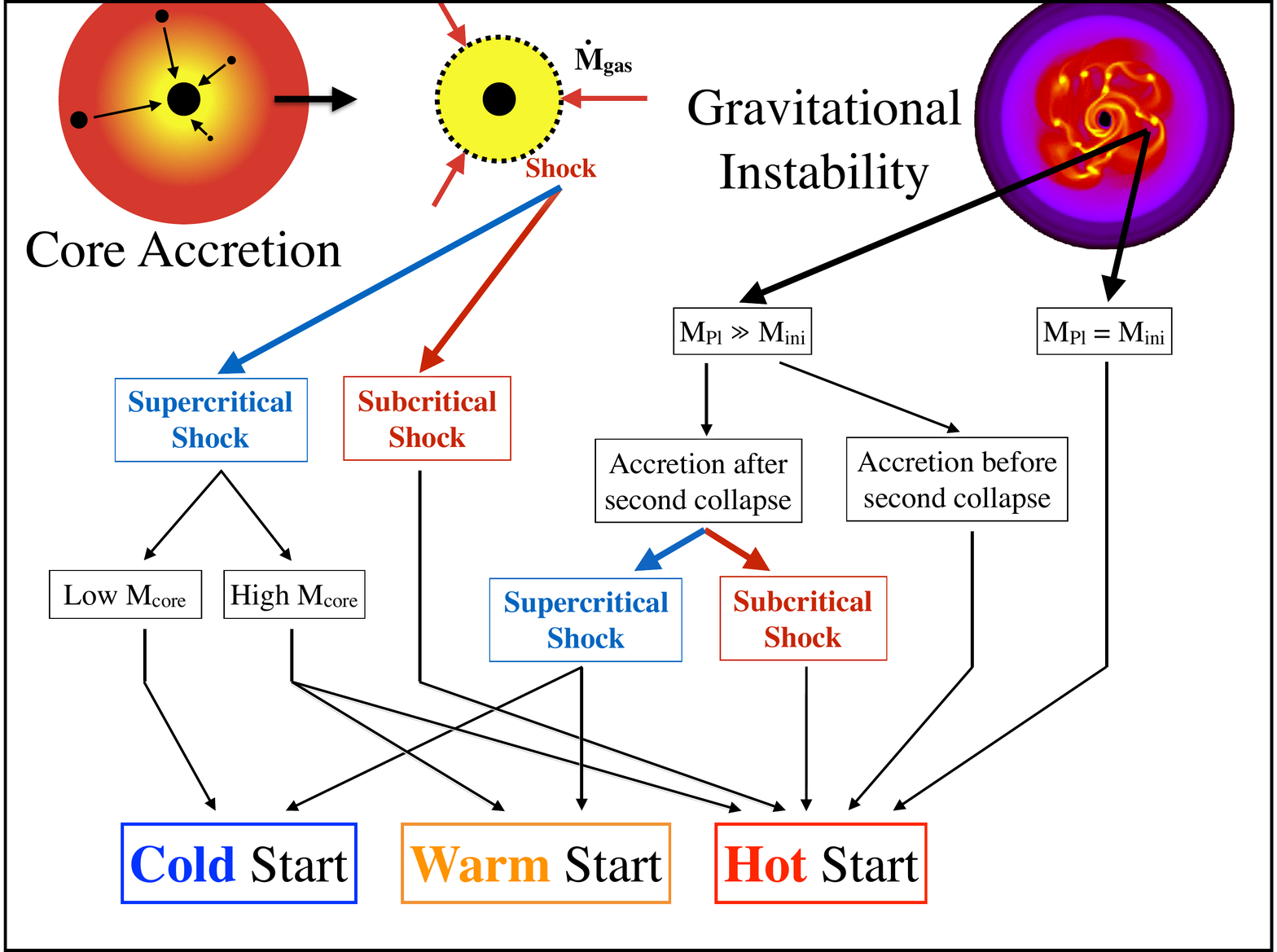}
\caption{Summary of the various pathways to giant planet formation
  leading to different post-formation entropies
  \citep{mordasinialibert2012b}.  The gas surface density plot in the
  upper-right corner illustrating gravitational instability is taken
  from \citet{boley2009}. $M_{\rm ini}$ denotes the initial mass of
  the clump formed by gravitational instability. Note that several
  pathways have to date not been simulated quantitatively.}
\label{fig:ini_cond_box}
\end{figure}

%-------------------------------------
The early evolution of gaseous planets depends crucially on the amount
of heat retained from the formation process. Depending on whether they
form with a low, average or high initial entropy (with a
correspondingly low, average and high luminosity, radius and
temperature), planets are said to have "cold", "warm" or "hot" start
initial conditions. A typical cold start post-formation entropy is
$\sim$ 8-9 $k_{\rm B}$/baryon ($k_{\rm B}$ denotes the Boltzmann
constant) whereas a typical hot start post-formation entropy is $\sim$
11-15 $k_{\rm B}$/baryon \citep[see, e.g.,][]
{marleyfortney2007,spiegel2011,mordasini2013,marleaucumming2013}.
Given the significant impact that such initial conditions can have on
the early internal evolution of giant planets, we first review the
expected range of initial entropies for the two mainstream scenarios
of giant planet formation: core accretion and disk gravitational
instability. This is also summarized in Figure~\ref{fig:ini_cond_box},
but it should be noted that several assumptions have not been
validated yet with quantitative simulations, especially in the
gravitational instability scenario.
% FFFFFFFFFFFFF
% FFFFFFFFFFFFF
\\
\par
{\bf (i) Core accretion --} Traditionally, planet formation via core
accretion is thought to produce gaseous planets with a low initial
entropy \citep{fortneymarley2005b,marleyfortney2007}. The reason for
this is the assumption that the accretion shock produced on the
planet's surface by the infalling gas is radiatively efficient (or
``super-critical''). For such a shock structure (found, e.g., for gas
accretion on the first stellar core, \citealt{commerconaudit2011}),
the gravitational potential energy released at the shock is fully
radiated away and none of it is incorporated in the convective
interior of the planet. By analogy with the stellar case
\citep[e.g.,][]{hartmanncassen1997}, planets undergoing such ``cold''
accretion will start their evolution with a low initial entropy and
low internal temperature, dubbed ``cold start''.

However, the assumption that the shock is radiatively efficient is
quite uncertain. If, on the contrary, the accretion shock is
radiatively inefficient (or ``sub-critical''), that is if the energy
released at the shock is not radiated away (as found for gas accretion
on the second stellar core, see \citealp{vaytetchabrier2013}), planets
will have high post-formation entropies, actually similar to those
assumed in classical purely evolutionary simulations with a hot start
\citep{mordasinialibert2012b}. The intermediate case where the
accretion shock is neither radiatively efficient nor inefficient will
lead to warm starts \citep{spiegelburrows2012}.  Furthermore, and
somewhat surprisingly, the formation of a massive solid core increases
the post-formation entropy \citep{mordasini2013,bodenheimer2013}. The
requirement for this is an increased solid surface density
($\Sigma_{\rm s}$) in the protoplanetary disk, which results in a
higher solid accretion luminosity within the planet and a higher final
core mass.  Planets forming in disks with a high $\Sigma_{\rm s}$ will
go into runaway gas accretion earlier, as the core also forms more
rapidly.  Therefore, gas is accreted onto a protoplanet with a larger
radius, as the planet has had less time to cool and is more strongly
stabilized against contraction owing to the increased planetesimal
accretion luminosity. A larger radius means a weaker shock, therefore
less gravitational potential energy released by the gas before it is
accreted onto the planet. The gas is then incorporated into the planet
at a larger entropy, also leading to warm starts
\citep{mordasini2013,bodenheimer2013}.
\\
\par
{\bf (ii) Disk gravitational instability --} Another pathway to the
formation of gaseous planets is gravitational instability (GI) in the
outer parts of protoplanetary disks \citep[typically beyond 50 to 100
AU; see, e.g.,][]{Boley10}. The possibility that GI occurs, the
location where fragmentation sets in, and the subsequent internal and
orbital evolutions of the clumps depend on the disk's mass, its
structure and, critically, its cooling time scale (see, e.g., the
reviews by \citealp{Helled_etal14} and \citealp{KratterLodato16}).
Analytical arguments based on the Toomre \citep{toomre1981} and
cooling criteria \citep{gammie2001} show that at smaller distances
from the star, in a disk that is dense enough to be gravitationally
unstable, it is not possible to cool rapidly enough to fragment, so
that the two criteria cannot be simultaneously fulfilled
\citep{rafikov2005}.  In the outer parts of the disk, gravitational
instability could instead occur early in the disk's evolution when it
is still massive and accreting rapidly from the protostellar cloud
\citep{dangelodurisen2010}. But it is then unclear if bound clumps
forming at this stage survive migration, evolve into planets, massive
brown dwarfs or stellar companions
\citep[e.g.,][]{StamatellosWhitworth2009}, or if they get tidally
disrupted \citep[e.g.,][]{Nayak10a,forganrice2013}.
  
Because in the case of a gravitational collapse the material is
thought to not process through a shock, GI should yield hot starts.
This has been found indeed in the numerical simulations of
\citet{galvagnihayfield2012}, who report an entropy of $\sim$ 15
$k_{\rm B}$/baryon. However, if the collapsed protoplanet continues to
accrete gas after it has fully collapsed (called ``second collapse'',
analogous to stellar formation, a second collapse denotes the collapse
after the molecular hydrogen has been dissociated) an accretion shock
can potentially develop.  The more mass is added through a
super-critical shock onto the initial GI unstable object, the lower
the final post-formation entropy of the planet
\citep{mordasinialibert2012b}. If, however, the initially unstable
object accretes its mass before the second collapse, it should rather
have a hot start as the large radius of the clump should prevent the
formation of strong shocks.
\\
\par
We point out that in ``classical" planetary evolution calculations,
which do not treat the formation phase, it is often simply assumed
that planets start with an arbitrarily high initial entropy (that is,
a hot start; \citealp[see, e.g.,][]{burrows1997,baraffechabrier2003}).
Because the Kelvin-Helmholtz (KH) timescale is short for large
entropies, the initial condition is quickly erased such that after
$\sim 10^7-10^9$ years for 1 and 10 $M_{\rm J}$ planets, respectively,
the luminosities in the cold and hot start cases converge to the same
evolutionary tracks \citep[see, e.g.,][]
{marleyfortney2007,spiegel2011,marleaucumming2013}.  Therefore, the
choice of an arbitrarily high initial entropy does not affect
predictions on the planet appearance at ages $>$ $10^7-10^9$
years. But it will do so for younger planets.  This is particularly
important for directly imaged planets in young stellar systems, for
which the deduction of the planet's mass based on the planet's
luminosity depends on the underlying assumptions of the utilized
planetary evolutionary tracks.

% FFFFFFFFFFFFF
\begin{figure}[t]
\centering
\includegraphics[width=0.8\textwidth]{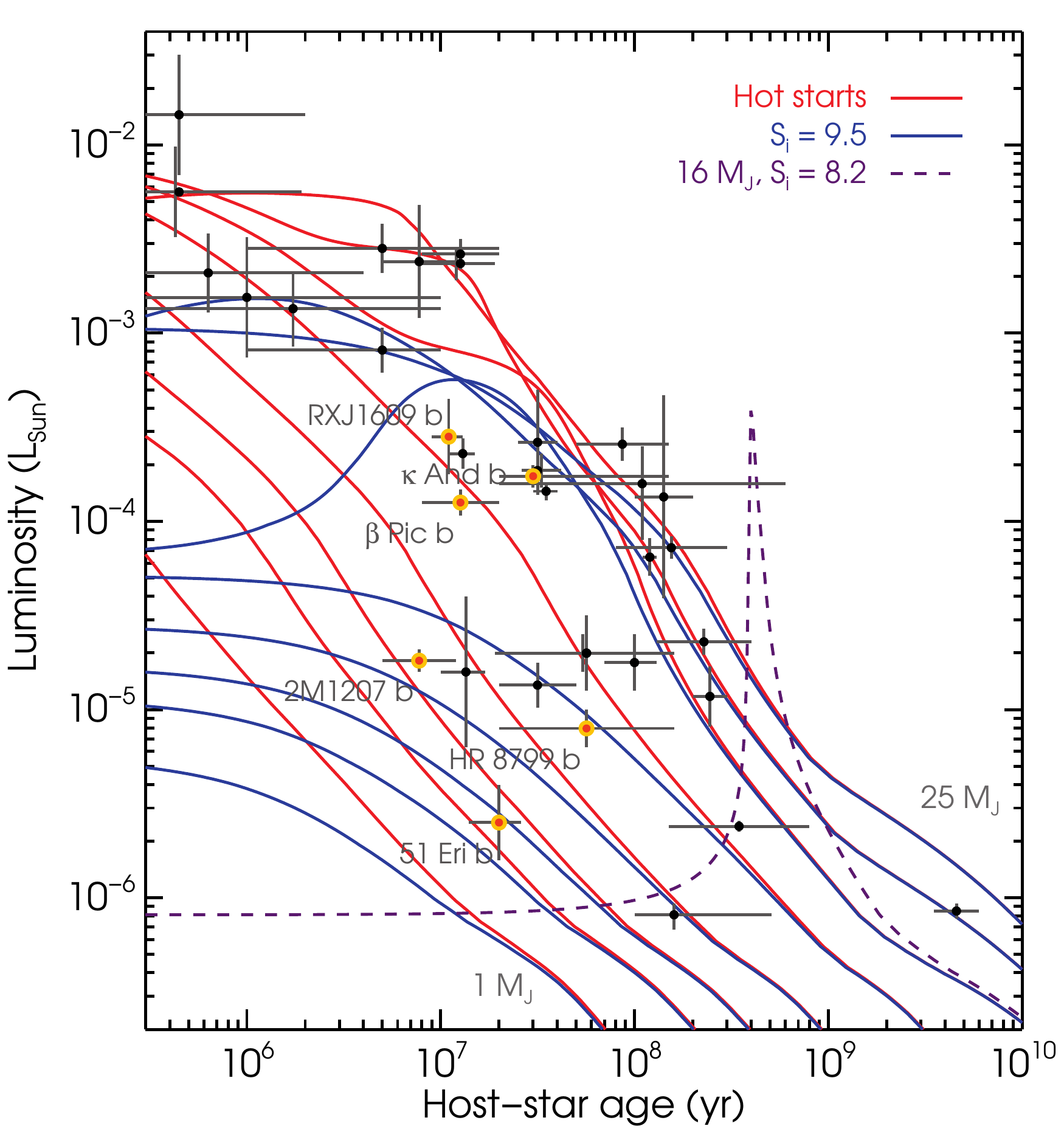}
\caption{Luminosity as a function of time from theoretical models
  \citep{marleaucumming2013} compared with directly imaged objects
  with a hot-start mass below 25 Jovian masses \citep[data compiled
  by][]{neuhauserschmidt2012}. The lines show cooling curves for $M$ =
  1, 2, 3, 5, 10, 15, 20 and 25 $M_{\rm J}$ (bottom to top) assuming a
  cold start with a post-formation entropy of 9.5 $k_{\rm B}$/baryon
  (blue lines) or a hot start (red lines). Figure updated from
  \citet{marleaucumming2013}.}
\label{fig:ini_cond}
\end{figure}
% FFFFFFFFFFFFF
A comparison between hot and cold start calculations is illustrated in
Figure~\ref{fig:ini_cond} (updated from
\citet{marleaucumming2013}). It shows the time evolution of the
luminosity of planets with a mass of 1, 2, 3, 5, 10, 15, 20 and 25
$M_{\rm J}$ for cold (blue curves) and hot initial conditions (red
curves).  Observational data of directly imaged planets are
overplotted (updated from \citealt{neuhauserschmidt2012}).  One can
see that for the specific evolutionary tracks shown in the figure,
cold and hot start tracks converge after $\sim$ 10$^8$ years. The mass
degeneracy for a given age and luminosity is obvious when looking at
the figure: at an age of $\sim 4 $ Myrs, a cold-start planet of 10
$M_{\rm J}$ and a hot-start planet of 3 $M_{\rm J}$ have the same
luminosity. For cooler starts than assumed in the figure, the time
after which the evolutionary tracks of cold and hot starts converge
becomes longer and the difference in masses for a given age and
luminosity increases.

%-------------------------------------
\subsection{Deuterium Burning in Massive Planets}
\label{subsect:dburn}
% -------------------------------------
Traditionally, gaseous objects which are able to fuse deuterium, but
which are not massive enough to burn hydrogen on a stable main
sequence, are called ``brown dwarfs'' \citep[see, e.g.,][]{boss2007}.
They were first theorized to exist in the 1960s
\citep{kumar1963,hayashi1963}. The distinction between massive gaseous
planets and brown dwarfs has been based on the hypothesized formation
mode, with formation in a protoplanetary disk for giant planets, and
gravoturbulent fragmentation of a molecular clump (i.e., a star-like
formation mode) for brown dwarfs \citep{chabrierjohansen2014}.  In
principle, it is well possible that, in massive protoplanetary disks,
planets form with masses in excess of 13 Jupiter masses via core
accretion \citep{mordasinialibert2009}, and it has been shown that
such planets are able to burn deuterium in the layers on top of their
solid core \citep{baraffe2008}.

Classical studies in the brown dwarf regime include
\citet{BurrowsLiebert1993,saumon1996,chabrier2000,burrows2001,baraffechabrier2003}.
They put a mass limit for the onset of deuterium burning at roughly 13
$M_{\rm J}$. Deuterium burning in forming planets that harbor a solid
core has been studied systematically in \citealt{molliere2012} and
\citet{bodenheimer2013}, and the mass threshold for deuterium burning
is found to be $\sim$13 $M_{\rm J}$ as well. For planets, the exact
threshold value can vary with the helium content, the metallicity and
deuterium fraction \citep{spiegel2011, molliere2012}. It also varies
with the core mass \citep{molliere2012,bodenheimer2013}, since a more
massive core leads to a higher post-formation entropy (see
Sect.~\ref{sect:ini_cond}) and therefore a higher internal
temperature, which increases the deuterium burning rate
\citep{mordasini2013,bodenheimer2013}.

In general, deuterium burning proceeds via the reaction \beq {\rm H} +
{\rm D} \rightarrow {^3{\rm He}} + \gamma \ , \eeq with an energy
release of 5.494 MeV \citep{fowler1967}.  Since the hydrogen and
deuterium ions need to be brought close enough to overcome the Coulomb
repulsion and fuse, the reaction rate depends very strongly on the
temperature. It is limited by the fact that, at the central
temperatures in the objects of interest, only a small fraction of ions
have kinetic energies large enough to tunnel through the ions'
repulsive Coulomb barrier. Screening plays an important role too, as
the electrons that are present in the fully ionized interiors of the
planet effectively shield the ions from each other during their
approach, decreasing the kinetic energy needed for the tunneling
probability. As the objects are already considerably degenerate in
their interior, classical Debye-H\"{u}ckel shielding is not sufficient
for these objects and screening theories for degenerate material need
to be invoked \citep[see, e.g.,][]{dewitt1973,graboske1973}.

Deuterium burning can have a very strong dependence on
temperature. The deuterium burning rate ($\epsilon_{\rm D}$, energy
per unit mass and time) can be expressed as a power-law function of
temperature $T$ \citep[see, e.g.,][]{kippenhahnweigert1990}: \beq
\epsilon_{\rm D} \propto \rho X_{\rm D} X_{\rm H}T^{\nu} \eeq with
$\rho$ the gas density, $X_{\rm H}$ and $X_{\rm D}$ the mass fractions
of hydrogen and deuterium, respectively, and \beq \nu =
\frac{80}{3}\left(\frac{10^5 {\rm K}}{T}\right)^{1/3} - \frac{2}{3}.
\eeq The power-law exponent of the temperature thus depends itself on
the temperature. For deuterium burning in low-mass protostars, it is
often assumed that $\nu = 11.8$ \citep{stahlerpalla2005}, which
corresponds to temperatures of $\sim $10$^6$~K. The coolest deuterium
burning objects (with masses of $\sim$~13~$M_{\rm J}$) burn deuterium
at central temperatures of roughly 4$\times$10$^5$~K \citep[see,
e.g.,][]{molliere2012}, corresponding to $\nu \sim 16$.  Note,
however, that electron screening in these objects may somewhat
decrease $\nu$ \citep{kippenhahnweigert1990}.

The outcome of deuterium burning is quite different whether a hot
start or a cold start is assumed. Hot start objects, which might be
rather related to ``classical'' brown dwarfs, start their evolution
with a large radius and a high temperature (see
Sect.~\ref{sect:ini_cond}), and then contract slowly with time,
starting the deuterium burning process once their central density is
high enough. The main effect of deuterium burning is to slow down the
contraction. Cold start objects, however, have quite low
post-formation entropies, resulting in small radii and small internal
temperatures (see Sect.~\ref{sect:ini_cond}).  The fact that their
internal temperature is smaller than in hot start objects, while
having smaller radii, is due to the partial degeneracy of their
interior.  When cold start planets reach internal temperatures that
are high enough for the onset of deuterium burning, they have much
higher densities than hot start objects. The effect of deuterium
burning for cold start planets is not a decreased contraction speed,
but a re-inflation to larger radii.  The increase in planetary
luminosity for these objects therefore does not exclusively stem from
the deuterium burning itself, but also from the later re-contraction
of the planets inflated by deuterium burning.

In general, the deuterium number fraction in the hydrogen and helium
dominated atmospheres is quite low, $N_{\rm D}/N_{\rm H} \sim 2\times
10^{-5}$ \citep{prodanovic2010}. Therefore, the phase of deuterium
burning is rather short, between 1 and 100 Myrs, depending on the mass
and formation mode (cold/hot).

% FFFFFFFFFFFFF
\begin{figure}[t]
\centering
\includegraphics[width=1.\textwidth]{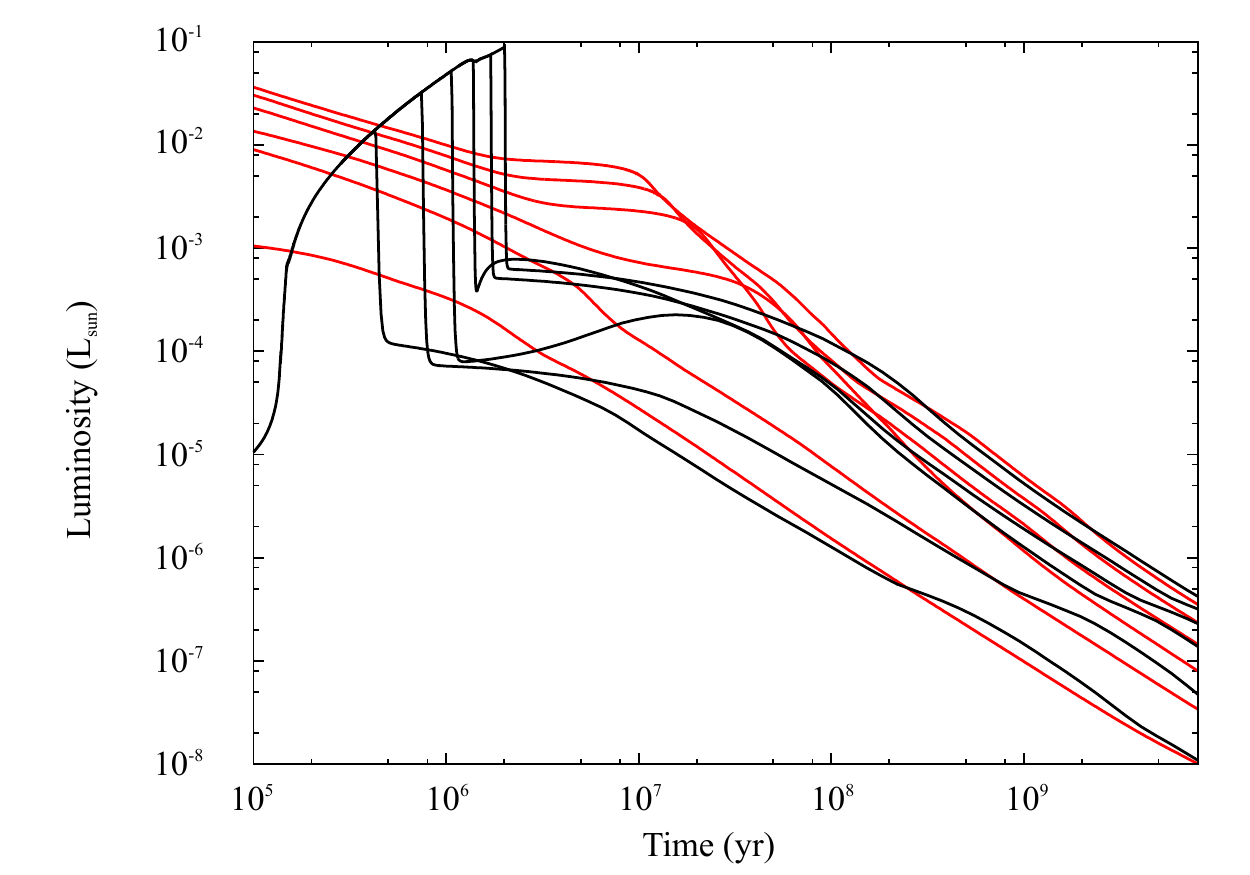}
\caption{Luminosities of cold and hot start objects with a mass of 5,
  10, 15, 20, 25 and 30 $M_{\rm J}$ (from bottom to top). The red
  solid curves show evolutionary calculations with a high initial
  entropy (hot start), taken from \citet{burrows1997}. The black solid
  curves show combined formation and evolutionary calculations of
  objects formed by core accretion with a low post-formation entropy
  (cold accretion). The high luminosities at around 1 Myr are due to
  the gas accretion shock, they rapidly fall as gas accretion is
  (artificially) terminated \citep{molliere2012}. }
\label{fig:dburning}
\end{figure}
% FFFFFFFFFFFFF
Illustrative luminosity tracks for hot start and cold accretion
objects are shown in Figure~\ref{fig:dburning}.  The hot start tracks
are taken from \citet{burrows1997}, while the cold start formation and
evolution calculations for core accretion planets use the model of
\citet{molliere2012}. The objects shown in the figure have a mass
ranging from 5 to 30 $M_{\rm J}$. Note that the low-entropy tracks in
Figure~\ref{fig:ini_cond} do not agree with those in
Figure~\ref{fig:dburning}: Figure~\ref{fig:ini_cond} assumes
arbitrarily low post-formation entropies for cold start objects,
whereas in Figure~\ref{fig:dburning} the low initial post formation
entropies are determined self-consistently from the formation
process. The point of Figure~\ref{fig:dburning} is that even objects
formed by core accretion produce deuterium burning rates similar (but
not identical) to those obtained by classical calculations which
normally neglect the formation process, while the point of
Figure~\ref{fig:ini_cond} is to show the differences in evolution
between high and low entropy objects and how long it takes for both
branches to converge. The formation is neither considered nor shown to
focus on the effects of evolution.

The effect of the deuterium burning delaying the hot start's
luminosity decrease for masses $>$ 13 $M_{\rm J}$ can clearly be seen
in Figure~\ref{fig:dburning}, as well as the increase in luminosity
for cold start objects (in order to compare the hot and cold start
calculations more easily, the cold start calculations were shifted in
time by roughly 2 Myrs). For the cold start objects with masses $>$ 20
$M_{\rm J}$, the inflation of the planetary radius happens already
during the formation phase of the planet, such that the increase in
luminosity cannot be seen in the figure as it lies within the period
where the powerful accretion shock luminosity dominates ($<$ 3 Myrs in
the figure). The accretion shock luminosity decreases a bit because of
deuterium burning as the infalling gas is added at larger radii.

%-------------------------------------
\subsection{Mass-Radius Relation}
\label{ssec:mrrelation}
%-------------------------------------
A diagram that is of particular interest for planet formation and
evolution theories is the planetary mass-radius diagram, as it
contains information about the internal structure of (extrasolar)
planets \citep[e.g.,][]{guillot1999,chabrierbaraffe2009}. It
constrains the processes that determine the structure during the
formation period (solid/gas accretion, orbital migration) as well as
during the evolutionary phase (like atmospheric opacities, energy
transport mechanisms, atmospheric escape, outgassing, or radius
inflation).

\begin{figure}[t]
\centering
\includegraphics[width=1.\textwidth]{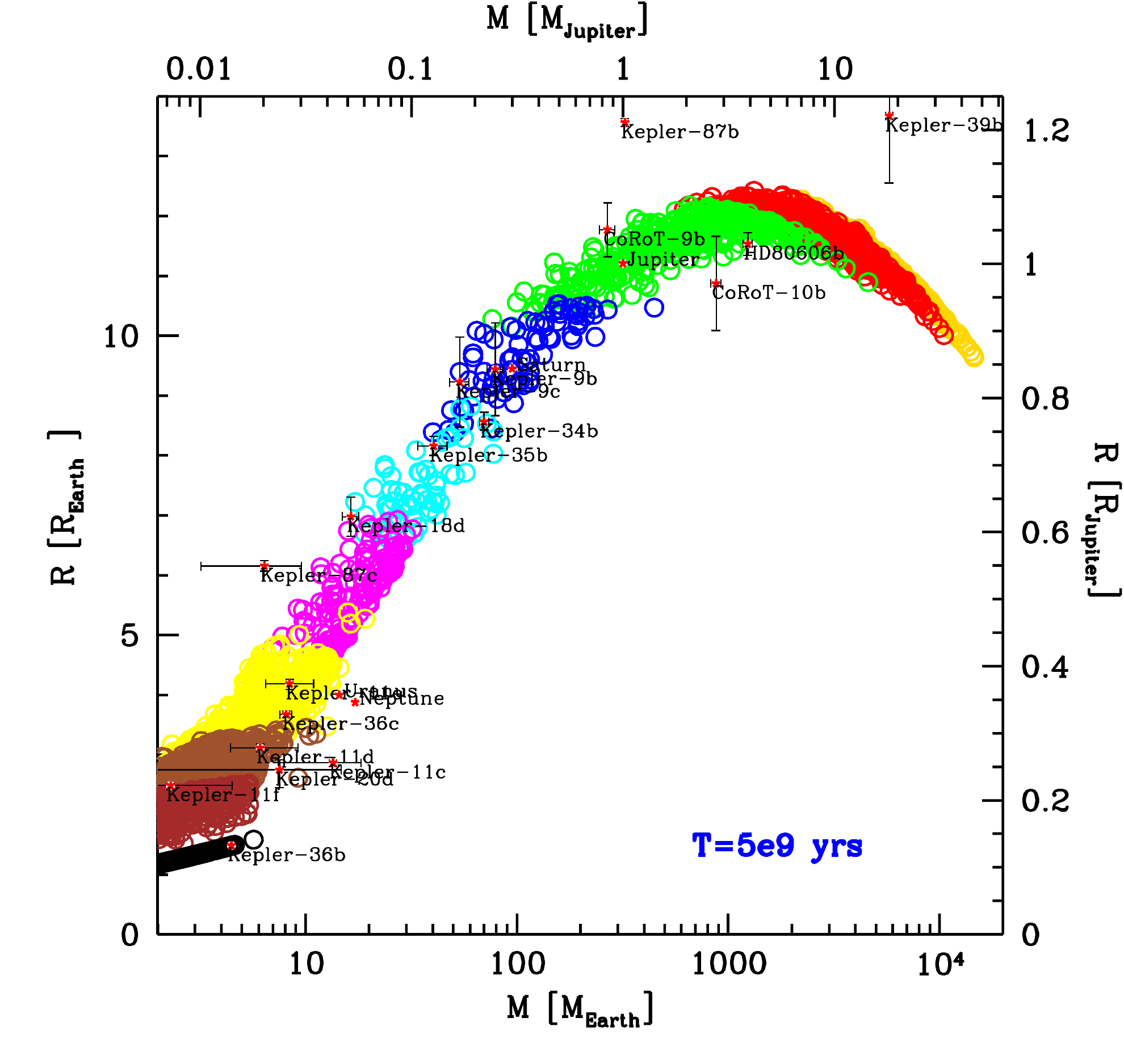}
\caption{Mass-radius diagram of observed and synthetic planets. The
  synthetic planets have an age of 5 Gyr and a semi-major axis $a$
  between 0.1 and 1 AU. Observed planets with $a>0.1$ AU are
  included. The colors give the mass fraction of H/He ranging from
  solid planets without H/He (black) over ocher ($<1$\%) to dark
  orange ($>99$\%).}
\label{fig:mr}
\end{figure}

Figure~\ref{fig:mr} compares the mass-radius relation obtained with a
planet population synthesis calculation at an age of 5 Gyr with
observational data (updated from \citealt{mordasinialibert2012c}). The
metallicity (core mass fraction) of the synthetic planets is
color-coded in the figure. The chemical composition of the heavy
elements depends on the planet's formation track. Heavy elements
accreted beyond the iceline in the parent protoplanetary disk are
assumed to consist of 50\% water ice and 50\% "rocky" material, while
material accreted inside the iceline is fully "rocky", which is itself
assumed to be made of 2/3 silicates and 1/3 iron. Such mass fractions
are approximately expected for the condensation of solar composition
gas \citep{lodders2003,santosadibekyan2015}.  Although the observed
exoplanets shown in Figure~\ref{fig:mr} are not at an exact age of 5
Gyr the observational errors on the host star's age often allow for
ages of about 5 Gyr.  Furthermore, the planetary radius evolution in
the Gyr time range is slow, such that the 5 Gyr snapshot should
represent planets of various ages within the Gyr regime reasonably
well.

The population synthesis model used in Figure~\ref{fig:mr}
self-consistently combines a planet formation model based on the core
accretion paradigm \citep{alibertmordasini2005} with the long-term
thermodynamical evolution of the planets after the dissipation of the
protoplanetary disk \citep{mordasinialibert2012b}. The simulation
includes the effect of atmospheric escape (see
Sect.~\ref{sect:enveevap_mrho_sol_gas}) but does not model radius
inflation (see Sect.~\ref{sect:rad_bloat}). To avoid planets which
could be significantly inflated \citep{demoryseager2011a}, the minimum
semi-major axis of the synthetic and observed planets included in the
figure is 0.1 AU. Mechanisms that could lead to inflated close-in
exoplanets will be discussed in Sect.~\ref{sect:rad_bloat}.

% FFFFFFFFFFFFF
% FFFFFFFFFFFFF
In general the synthetic mass-radius diagram has a specific shape that
is characterized by an increase in radius with mass, which levels off
and turns over at roughly 4 M$_{\rm J}$.  This maximum is due to an
increase in the compressibility of the matter as the electron
degeneracy increases. Furthermore, one sees that neither the
upper-left corner in Figure~\ref{fig:mr} (i.e., large low-mass
planets) nor its lower-right corner (i.e., small massive planets) seem
to be populated by synthetic or observed planets.  Synthetic planets
with a low metal fraction have larger radii and masses than
high-metallicity synthetic planets. At a given mass, the lower the
radius, the more enriched is a synthetic planet, while planets of
fixed enrichment have larger radii when their mass is higher, leading
to diagonally colored bands in the synthetic mass-radius diagram.

% FFFFFFFFFFFFF
\begin{figure}[t]
\centering
\includegraphics[width=0.8\textwidth]{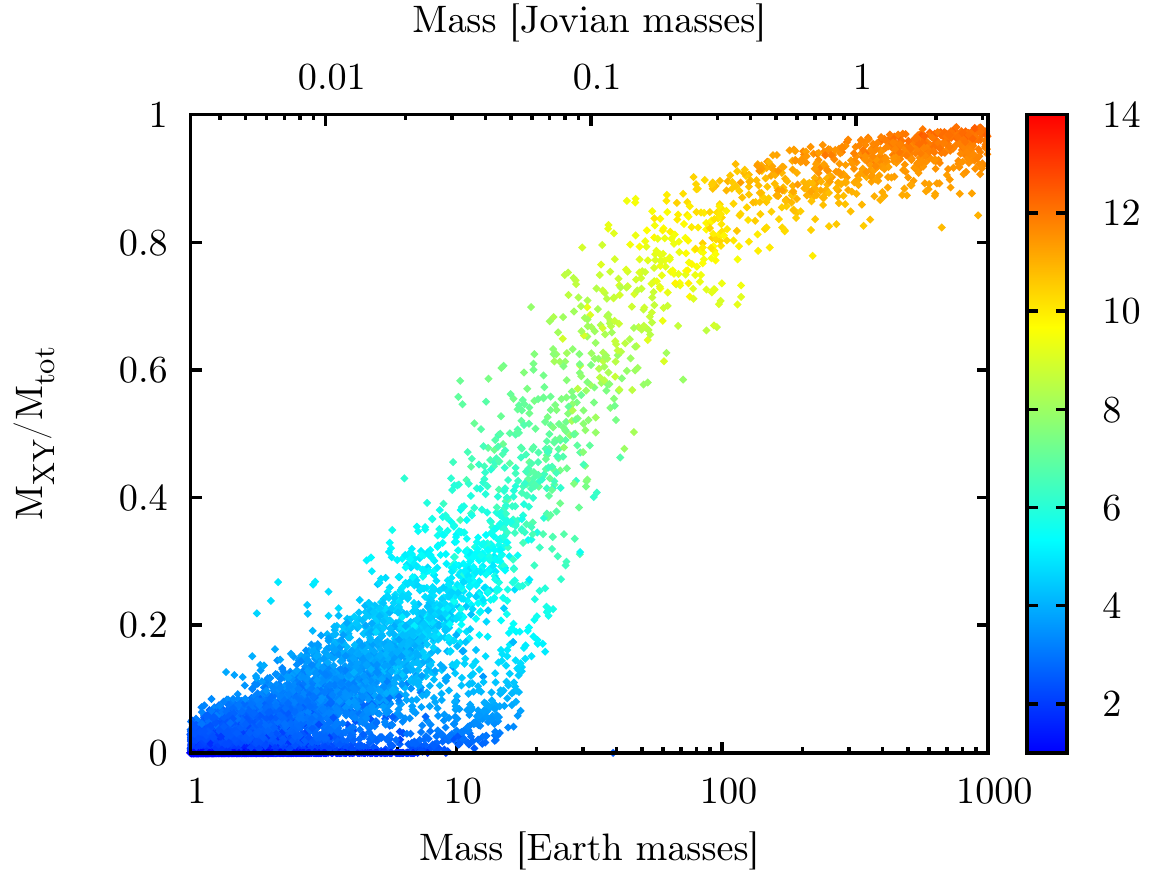}
\caption{Mass fraction of H/He relative to the total mass of a planet
  as a function of its total mass. The color-code gives the planetary
  radius in Earth radii. The larger the planet mass, the higher the
  mass fraction of the gas envelope. The change in slope at about 100
  $M_{\oplus}$ results from the transition between the
  Kelvin-Helmholtz contraction limited regime and the disk-limited gas
  accretion regime. This figure can be compared with observational
  constraints (see, e.g., Figure~9 in \citealt{lopezfortney2013b}).}
\label{fig:mmxym}
\end{figure}
% FFFFFFFFFFFFF
The shape of the mass-radius diagram is a consequence of the formation
process, in particular the resulting core and H/He envelope masses, as
well as material properties. The fraction of H/He envelope mass as a
function of the planet's total mass is displayed in
Figure~\ref{fig:mmxym}. As the population synthesis is calculated
under the core accretion paradigm, the planets start as low-mass solid
cores, which can bind only tenuous gaseous envelopes due to long KH
cooling timescales of their protoplanetary envelope during the nebular
phase \citep{ikomanakazawa2000,leechiang2015}. Additionally, if the
protoplanets are close to their star during formation, they will not
accrete large amounts of gas due to the high local disk temperature
\citep{ikomahori2012}. Whatever gaseous envelope they accrete during
formation can then be reduced again by envelope evaporation due to the
stellar XUV insolation during evolution
\citep[e.g.,][]{lammerselsis2003,owenjackson2012,lopezfortney2013,jin2014}.
Therefore, the lowest mass objects will have the highest heavy element
content (essentially they are solid rocky or icy planets). As rocky or
icy planets are only weakly compressible, their radius increases as
their mass increases.

More massive planets can bind a larger gaseous envelope on top of
their solid core. As they grow in mass, the amount of planetesimals
which can be accreted increases as well due to the expansion of the
feeding zone, but the net effect is still a decrease in the planetary
metallicity (heavy element mass fraction) with mass $M$ roughly as
$M^{-0.7}$ in the giant planet regime \citep{mordasiniklahr2014}, in
agreement with observational data \citep{millerfortney2011}.

Although the compressibility of gas is much higher than of solids, the
planets still grow in radius for a given metallicity if the mass is
increased.  For ever higher solid core masses, the growing planets
will eventually try to accrete and bind more gas from the
protoplanetary disk than is locally available due to their very short
KH timescales. This leads to a quasi-static collapse of their gaseous
envelope and the so-called stage of runaway gas accretion
\citep{bodenheimerhubickyj2000}. In this phase, the growing planets
will accrete a large fraction of the gas delivered to their location
by the viscous evolution of the protoplanetary disk
\citep{lubowseibert1999}.  The disk's lifetime and gas supply at the
location of the planet then become the sole determining factors of the
planets' final mass. The transition from the regime where the planet's
KH-contraction controls gas accretion, to the one where the disk
limits it, is visible in Figure~\ref{fig:mmxym} by a change in slope
at about 100 $M_{\oplus}$. This figure can be compared to
observational data where the H/He mass fraction of actual exoplanets
is derived from internal structure modeling (Figure~9 in
\citealt{lopezfortney2013b}).

As it is inherently impossible to form low-mass gaseous planets
without a core in the core accretion paradigm, no very large low-mass
planets exist in the synthetic mass-radius diagram (nor actually in
the observed diagram). Such planets would populate the upper-left
corner in Figure~\ref{fig:mr} (and Figure~\ref{fig:mmxym}). The
presence of a massive core always leads to comparatively smaller radii
for low-mass planets, even if there is a gaseous envelope on top of
the core. On the other hand, massive solid planets will always undergo
runaway gas accretion in the disk, such that it is impossible to form
small (i.e., bare rocky or icy) but massive planets in the core
accretion picture. This explains why the lower-right corner of
Figure~\ref{fig:mr} (and Figure~\ref{fig:mmxym}) also remains empty.

We see that the majority of the observed exoplanets fall within the
mass-radius diagram obtained with the population synthesis. Kepler-87b
and Kepler-39b are notable exceptions, being larger than the predicted
synthetic radii of any given mass. The structure of these planets
remains to be explained \citep{bouchybonomo2011}.  The large radius of
Kepler-39b could be caused by rapid rotation, its rotation period
being estimated at only 1.6 $\pm$ 0.4 hours \citep{zhuhuang2014}.  It
is interesting to note that some objects similar to Kepler-39b in
terms of mass, but at even higher insolation, follow the prediction of
population syntheses much better \citep[like CoRoT 3b, which has a
mass of 21.7 $\pm$ 1.0 M$_{\rm J}$ and a radius of 1.01 $\pm$ 0.07
R$_{\rm J}$ ,see][]{moutoudeleuil2013,deleuildeeg2008}.

The fact that some of the observed exoplanets with a mass comparable
to that of Neptune seem to be smaller than predicted by the model
needs to be explained as well. There are two potential explanations.
A first possibility is a higher opacity in the protoplanetary envelope
during the formation phase. The accretion of H/He during formation is
regulated by the envelope's ability to cool, which is in turn mostly
controlled by the opacity caused by small grains suspended in the
planet's upper radiative zone. If the opacity is higher than assumed
in this population synthesis (it is taken here equal to the ISM
opacity multiplied by a factor of 0.003,
\citealt{mordasiniklahr2014}), the planets would have smaller radii at
a given mass due to a lower H/He fraction. This shows how the
planetary mass-radius relation can be used to constrain the
microphysical models of the grain dynamics during formation like
coagulation, settling and evaporation
\citep{podolak2003,ormel2014,mordasini2014}. A second possibility is a
more efficient envelope evaporation during evolution. If the loss of
the primordial H/He envelopes is actually more efficient than in the
simple evaporation model taken here (e.g., because of different
evaporation regimes applying to planets of different types,
\citealt{owenalvarez2015}), this would also lead to smaller radius at
a given mass. These two scenarios could potentially be disentangled by
their different dependence on the planet's orbital distance, the
planet's atmospheric composition, or on time. But in order to see such
features, a high number of well-characterized planets is likely
necessary, highlighting the importance of future missions like CHEOPS
\citep{broegfortier2013} or PLATO \citep{rauercatala2014}.

%-------------------------------------
\subsection{The Formation of Close-in Low-Mass Planets, Envelope
  Evaporation, and the Planetary Mass-Density Diagram}
\label{sect:enveevap_mrho_sol_gas}
%-------------------------------------
Close-in low-mass planets are extremely common around Sun-like stars
with about 45-55 \% of FGK stars having a planet with $M\lesssim 30
M_{\oplus}$ and a period $P<100$ days, a result from high-precision RV
surveys \citep{udrymayor2008,mayormarmier2011} that was subsequently
confirmed and extended by the {\it Kepler} survey
(\citealt{boruckikoch2011c}).

Despite their high frequency, the formation mechanism of this class of
planets (without analogue in the Solar system) is currently debated in
the literature \citep[e.g.,][and
Sect.~\ref{ssec:multi}]{hansenmurray2012,chianglaughlin2013,idalin2013,raymondcossou2014,schlichting2014,handsalexander2014,chatterjeetan2014,ogiharamorbidelli2015,inamdaraschlichting2015}. Some
models propose an in situ formation at the current location of the
planets close to the star, other models propose that the planets have
originally formed further away, potentially in the colder regions of
the protoplanetary disk beyond the water ice line, while some other
models finally are of a hybrid type where at least some migration of
the planets themselves or their building blocks has occurred. As
summarized in Table~\ref{tab:closeinlowmass}, these models differ in
their predictions not only regarding the architecture of the planetary
systems (e.g., mean motion resonances, concentration towards the inner
edge, relative size or mass of the planets), but also in their
predicted bulk composition (ice mass fraction in the planetary core,
H/He mass fraction).
% TTTTTTTTTTTTTT
\begin{table}
  \caption{Predictions of three scenarios for the formation of close-in low-mass planets 
    in terms of the resulting core composition of the planets, whether planets 
    attain a primordial H/He envelope, are expected to be in mean motion resonances, 
    and the disk mass necessary to form the planets. The term isolation refers to the 
    situation where the planetesimals' orbits are sufficiently circularized by the protoplanetary 
    gas disk so that they can never grow enough in mass by collisions and cannot attract a 
    large envelope of gas during their formation.}
\label{tab:closeinlowmass}
\centering
\begin{tabular}{|l||*{3}{c|}}
  \hline
  \backslashbox{Quantity}{Scenario} &  \pbox{0.25\hsize}{Strictly in situ/\\no migration}  & \pbox{0.25\hsize}{Weak migration,\\then final assembly} & \pbox{0.25\hsize}{Assembly at $a\gtrsim a_{\rm ice}$,\\large scale migration}  \\[0.2cm] 
  \hline\hline
  Core composition & Rocky & \pbox{0.25\hsize}{Rocky to icy with\\potential gradient}   & Icy \\[0.2cm]  \hline
  Primordial H/He  & \pbox{0.25\hsize}{\vspace{1mm} Yes (w/o isolation)\\No (w isolation)} &\pbox{0.25\hsize}{\vspace{1mm} Yes (w/o isolation)\\No (w isolation)} & Yes \\[0.2cm]  
  \hline
  Mean-Motion Resonance & No & No/Yes & Yes/No (see Sect.~\ref{ssec:multi}) \\ 
  \hline
  \pbox{0.25\hsize}{\vspace{1mm} Necessary disk mass}& High & Intermediate & Low \\[0.1cm]  \hline
\end{tabular}
\end{table}
% TTTTTTTTTTTTTT
The table makes clear that knowing the bulk composition (H/He content;
rocky, icy, mixed core composition) and also the atmospheric
composition would be of very high interest to disentangle the
different formation mechanisms. The different bulk compositions
express themselves in different mean planetary densities, even if
degeneracies unfortunately prevent a unique conversion of the observed
mean density into an underlying bulk composition, at least for some
parameters \citep[e.g.,][]{adamsseager2008,dornkhan2015}.

Figure~\ref{fig:mrho} shows the planetary mass-mean density diagram
obtained in a population synthesis calculation based on the core
accretion paradigm and assuming Sun-like stars (Swoboda et al., in
prep.). During the formation phase, the concurrent formation of 10
embryos was calculated using the model of \citet{alibertcarron2013},
while during the subsequent long-term evolution, the effects of
cooling and contraction as well as atmospheric escape were included
\citep{mordasinialibert2012b,jin2014}. Atmospheric escape is an
important process for low-mass close-in planets because (i) they are
more susceptible to it than giant planets due to their weaker
gravitational potential, and (ii) the loss of even a tenuous H/He
envelope strongly decreases their radius
\citep[e.g.,][]{lopezfortney2013}. Envelope evaporation has been
observed both for giant and low-mass planets
\citep{vidalmadjarlecavelier2008,ehrenreichbourrier2015}. The
mass-density diagram is interesting because it reveals more clearly
than the mass-radius diagram the imprint of several prominent physical
processes during planet formation and evolution \citep[for a
comparison with an observational $M-\bar{\rho}$ diagram,
see][]{HatzesRauer15}. They can be identified with seven features in
Figure~\ref{fig:mrho}:

% FFFFFFFFFFFFF
\begin{figure}[t]
\centering
\includegraphics[width=1.\textwidth]{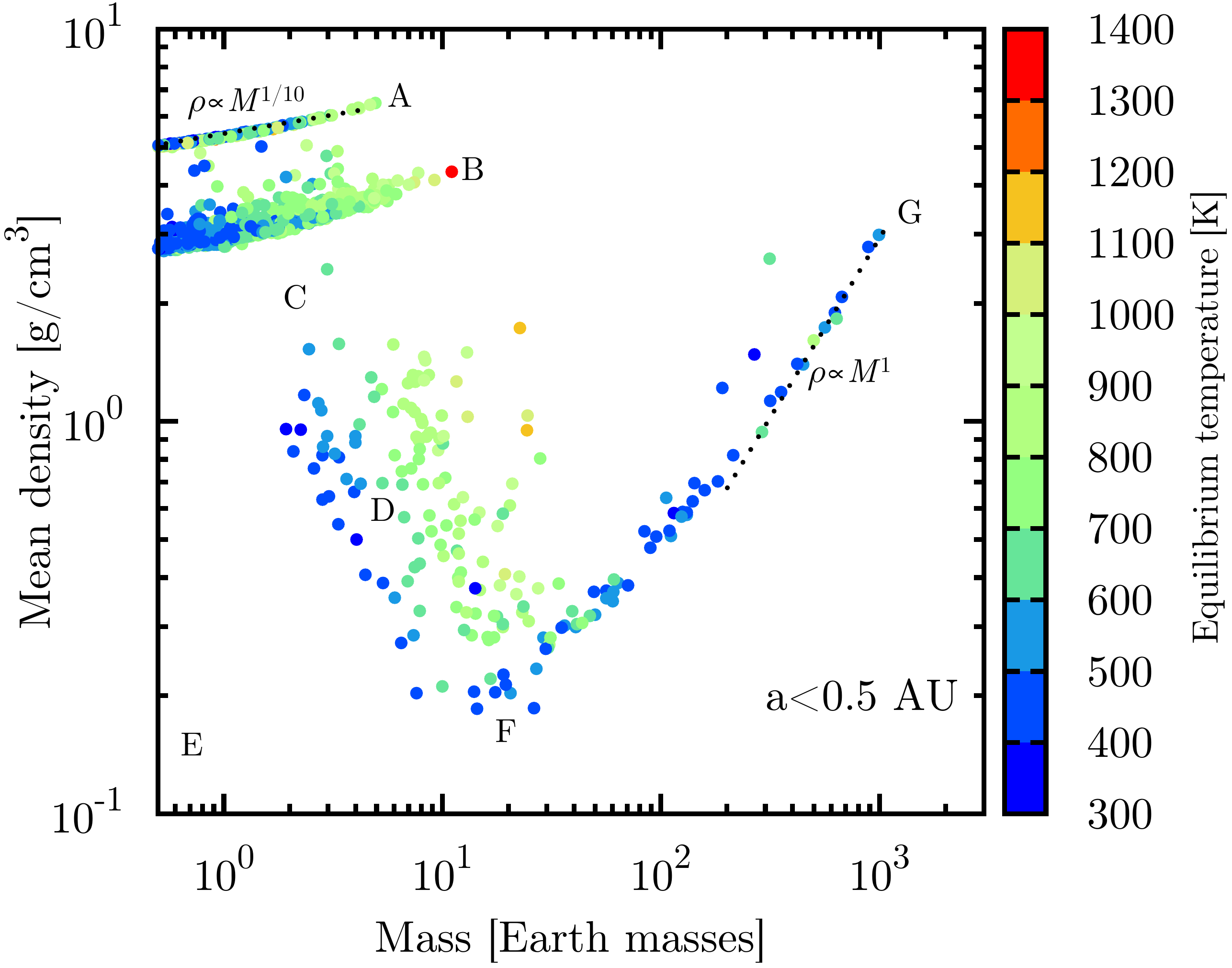}
\caption{Mean planetary density $\bar{\rho}$ as a function of mass $M$
  in a synthetic population around Sun-like stars at an age of 5
  Gyr. Planets have a semi-major axis $a$ $<$ 0.5 AU. The color bar
  gives the equilibrium temperature of the planets. Important features
  are: A) bare rocky planets; B) bare icy planets; C) the evaporation
  valley; D) low-mass core-dominated planets with H/He; E) a forbidden
  zone because of evaporation ($a$ $<$ 0.5 AU); F) the transition
  point to giant planets; G) the giant planet branch. The analytic
  scaling $\bar{\rho} \propto M^q$ for a polytropic index of $n=1/7$
  ($q=1/10)$ applying to terrestrial planets and $n=1$ ($q=1)$ for
  partially degenerate giant planets are overplotted by dotted lines.}
\label{fig:mrho}
\end{figure}
% FFFFFFFFFFFFF

\begin{description}
\item[\textbf{A: bare rocky planets.}] These are low-mass rocky
  planets which have either started without an H/He envelope (very
  low-mass planets) or which have lost it completely due to
  atmospheric escape. These planets have formed inside the iceline but
  have migrated to their current orbital distance. In the simulation,
  they fall on one single mass-density relation as a universal mass
  fraction of iron (1/3) was assumed in the synthesis, whereas in
  reality a certain spread is expected due to varying stellar
  compositions \citep{dornkhan2015} and to giant impacts which can
  partially remove the mantle \citep{benzanic2007}. The colors show
  that at smaller orbital distances (corresponding to higher
  equilibrium temperatures $T_{\rm eq}$), more massive planets have
  lost all their primordial H/He. This illustrates that the dividing
  line between solid and gaseous planets depends on orbital distance
  if it is only due to escape, which is however unlikely
  \citep{schlichtingsari2015,rogers2015}.
\item[\textbf{B: bare icy planets.}] These are low-mass planets formed
  outside the iceline which have migrated close to their host star and
  then have lost their H/He due to evaporation. In the syntheses of
  \citet{alibertcarron2013}, the planetary ice-mass fraction increases
  with increasing orbital distance and planetary mass, as more massive
  planets migrate faster in type I migration, and get more easily lost
  out of type I migration convergence zones (``traps'') due to
  saturation of the corotation torque \citep[e.g.,][and
  Sect.~\ref{ssec:corotation}]{BaruteauPP6,dittkristmordasini2014}. The
  mass-density relation is less narrow than for the bare rocky planets
  indicating a spread in the ice mass fraction, despite the fact that
  the increase in the planetesimals' surface density at the iceline (a
  factor of 2) was the same in all synthetic disks. This means that
  all planets accreting outside of the iceline start with an identical
  ice mass fraction in the core. The final spread is then a
  consequence of the accretion of rocky planetesimals during the
  migration through the inner disk (see \citealt{figueirapont2009} for
  the case of GJ436b).
\item[\textbf{C: evaporation valley.}] The evaporation valley
  separates planets without H/He from low-mass planets that still have
  retained some H/He \citep{owenwu2013,lopezfortney2013,jin2014}. It
  comes about as a small amount of H/He (just 0.1-1\% in mass) is
  sufficient to decrease $\bar{\rho}$ by a factor 2 or so, and because
  the timescale over which this last H/He is lost is short
  ($\sim10^{5}$ years) compared to the planet's lifetime. This means
  that it is not likely that at a given moment in time, like here 5
  Gyrs, many planets are observed during the process of losing the
  last H/He, leading to the depletion of points in the valley.
\item[\textbf{D: low-mass core-dominated planets with H/He.}] These
  are planets in the super-Earth mass range. They can have both icy or
  rocky cores depending on their formation history. A clear imprint of
  evaporation is seen: the hotter a planet, the larger its density at
  a given total mass as more primordial H/He is lost for closer-in
  planets.
\item[\textbf{E: forbidden zone because of evaporation.}] This lower
  left triangle remains empty as only planets with $a<0.5$ AU are
  included, and at such distances, low-mass, very low-density planets
  would undergo intense evaporation. Therefore, these planets have
  already moved to higher mean densities by 5 Gyr. This means that the
  parameter space is in principle not only three-dimensional ($M,
  \bar{\rho}, T_{\rm eq}$), but that time should be an extra dimension.
\item[\textbf{F: transition point to giant planets.}] The point of
  lowest mean density in the diagram can be defined as the transition point from
  solid-dominated to gas-dominated planets. It occurs at a mass of
  10-30 $M_{\oplus}$, which corresponds to the mass where rapid gas
  accretion starts in the core-accretion model
  \citep{pollackhubickyj1996}. In the planetary mass function, a break
  in the slope can be seen at such a mass in both the theoretical and
  observational data \citep{mordasinialibert2009,mayormarmier2011}.
\item[\textbf{G: giant planet branch.}]  With further increasing mass,
  planets become denser again as the self-gravitational compaction
  becomes stronger for these gas-dominated, more compressible
  planets. In this regime, closer-in planets have a lower density
  (even if this is not clearly visible in this specific synthesis due
  to the low number of giant planets inside of 0.5 AU), which is the
  opposite to the core-dominated planets with H/He. For giant plants,
  the change of the atmospheric boundary conditions for more strongly
  irradiated planets delays somewhat the contraction
  \citep{guillot2002}, resulting in lower densities at higher $T_{\rm
    eq}$. Additional radius inflation mechanisms (not included in this
  synthesis) tend to render this correlation even clearer
  \citep[e.g.,][]{LaughlinCrismani2011}.
\end{description}

Other features in the mass-mean density diagram, in particular the
mass scalings, can be understood by considering polytropic models for
planetary interiors for which the pressure $P$ depends on the density
$\rho$ as $P\propto\rho^{1+1/n}$, with $n$ the polytropic index
\citep{chandrasekhar1939,chabrierbaraffe2009}. For such models, the
mean density scales with mass as $\bar{\rho} \propto M^{2n/(3-n)}$.

In the giant planet regime the electrons are partially degenerate due
to the relatively low temperatures and high densities. But the
degeneracy is not complete so that both the quantum
electron-degeneracy pressure and the classical electrostatic
contribution from the non-degenerate ions matter. The polytropic index
is therefore $n=1$ which corresponds to a (mean) density that
increases linearly with mass ($\bar{\rho}\propto M$) or, equivalently,
to a radius that is independent of mass. Indeed, such a behavior is
seen in the numerical simulations, as shown by the right dotted line
close to Feature G (the giant planet branch). At lower masses, the
degeneracy decreases and the composition changes from gas-dominated to
solid-dominated planets. These planets are described by equations of
state which are much less compressible.

For completely incompressible matter ($n$=0), the mean density would
be independent of mass, corresponding to a horizontal line in
Figure~\ref{fig:mrho}. Such a regime is approached in the case of
low-mass solid planets. The small, but still non-zero compressibility
of rocks, ices, and iron however causes a weak increase of the density
with mass also in this regime (see Features A and B). For low-mass
solid planets, the mass-radius relation scales approximately as
$R\propto M^{3/10}$ \citep{grassetschneider2009}, corresponding to
$n=1/7$, and therefore $\bar{\rho} \propto M^{1/10}$ which is a weak
increase. Such scaling is shown by the left dotted line in
Figure~\ref{fig:mrho}. It approximately reproduces the results found
by solving directly the interior structure equations of bare rocky or
icy planets \citep{mordasinialibert2012c}.

%TTTTTTTTT
\begin{table}
  \caption{Power-law indices governing the structure of 
    various types of planets described by polytropic models (see text): $n=1/(\gamma-1)$ 
    is the polytropic index, $\gamma$ the adiabatic index, $p = (1-n)/(3-n)$ describes 
    the power-law dependence between radius and mass, $q =2n/(3-n)$ that between 
    mean density and mass, and $\chi_{\rm norm} = 1/\gamma$ that between mean density 
    and pressure.}
\label{tab:polytropes}
\centering
\begin{tabular}{|l||*{4}{c|}}
  \hline
  \backslashbox{Planet type}{Quantity} &  Polytropic index $n$  & \pbox{0.25\hsize}{$p=\partial\log R / \partial\log M$} & \pbox{0.25\hsize}{$q=\partial\log \bar{\rho} / \partial\log M$} & \pbox{0.25\hsize}{$\chi_{\rm norm}=\partial {\rm log}\bar{\rho}/\partial {\rm log}P$}\\[0.2cm] 
  \hline\hline
  Incompressible & 0 & 1/3  & 0 & 0 \\  \hline
  Terrestrial planets & 1/7 & 3/10 & 1/10 & 1/8 \\ \hline
  Giant planets  &  1 & 0 & 1 & 1/2  \\ \hline
  Low mass BDs & 6/5 & -1/9 & 4/3 & 6/11 \\ \hline
  High mass BDs & 3/2 & -1/3 & 2 & 3/5 \\ \hline
\end{tabular}
\end{table}
%TTTTTTTTT
The scaling relations for polytropic models are summarized in
Table~\ref{tab:polytropes}. The results are shown for planets that are
approximately incompressible (very low-mass planets), terrestrial
planets, partially degenerate giant planets with masses roughly
between 1 and 10 $M_{\rm J}$, more strongly degenerate high-mass giant
planets and low-mass brown dwarfs with masses from about 10 to 30
$M_{\rm J}$, and finally high-mass brown dwarfs which can be
approximated with a fully degenerate non-relativistic equation of
state (n=3/2). The table gives the polytropic index $n$, the power-law
exponent $p$ for the radius $R$ as function of mass $M$ ($R\propto
M^{p}$), which, from the polytropic Lane-Emden equation, is linked to
$n$ via $p=(1-n)/(3-n)$. The table also gives the power-law exponent
$q$ for the mean density as function of mass ($\bar{\rho}\propto
M^{q}$), given by $q=2n/(3-n)$, and finally the dimensionless
power-law compressibility $\chi_{\rm norm} =
\partial {\rm log}\bar{\rho}/\partial {\rm log}P$, i.e. $\chi_{\rm norm} =
n/(n + 1) = 1/\gamma$ ($\gamma$ denotes the adiabatic index). This is
a good measure of the compressibility because it shows that for small
values of $\chi_{\rm norm}$ the density will only increase weakly even
if the pressure increases strongly. For large values of $\chi_{\rm
  norm}$, small changes in the pressure can change the density
substantially. The quantity $\chi_{\rm norm}$ increases with
increasing mass. In reality, the transition from one regime to the
other is smooth, as can be seen in Figure~\ref{fig:mr}. In this figure
we see for example that, for low-mass giant planets, the radius
increases with mass indicating an effective $n<1$, followed by the
maximum at about 4 $M_{\rm J}$ where $n=1$, and finally a smooth
increase to even higher $n$ for more massive objects that only ends
when hydrogen burning sets in at about 80 Jovian masses.

%-------------------------------------
\subsection{Radius inflation of Hot Jupiters}
\label{sect:rad_bloat}
%-------------------------------------
Radius inflation is a distinctive feature of close-in giant planets
which was not expected theoretically and which is still not fully
understood. With a radius of about 1.38 $R_{\rm J}$, HD 209458 b, the
first detected transiting extrasolar planet
\citep{HenryMarcy2000,CharbonneauBrown2000}, is significantly larger
than expected from theoretical evolution models, even when the effects
of stellar irradiation are taken into account
\citep{guillot2002}. This can be seen in Figure~\ref{fig:mr} which
shows that standard cooling models do not predict the existence of
planets with radii larger than about 1.1 $R_{\rm J}$ (or up to 1.3
$R_{\rm J}$ if smaller orbital distances are included) at ages of a
few Gyrs. Yet, a significant fraction of hot Jupiters are larger than
above threshold values, with some extreme cases such as HAT-P-32 b
($R$=1.79$\pm$0.03 $R_{\rm J}$, \citealt{hartmanbakos2011}) and WASP
17 b ($R$=1.93$\pm$0.08 $R_{\rm J}$,
\citealt{AndersonHellier2010}). The latter has a mean density about 10
times smaller than Jupiter's.

At the time of writing, thanks to a fairly large sample of transiting
giant planets with a wide range of orbital distances, it is fairly
well established that there exists a correlation between inflation and
the intensity of stellar irradiation \citep{LaughlinCrismani2011}. It
is found that modestly irradiated planets do not have inflated radii
and that below a threshold value for the incident flux of about
2$\times$10$^{8}$ erg s$^{-1}$ cm$^{-2}$ (corresponding to a
solar-like star at an orbital distance of about 0.08 AU) radius
inflation ceases \citep{demoryseager2011a}.

The observation that radius inflation, also called radius {\it
  anomaly}, is correlated with the incident stellar flux may therefore
indicate that some of the incident stellar energy is deposited deep
inside the planets' interior, thereby delaying or even halting their
contraction.  For this to work, part of the stellar irradiation flux
must be converted into some other form of energy that is then
deposited in the deeper layers, as pure irradiation of the planet's
atmosphere by the star does not penetrate deep enough into the planet
to cause a large enough inflation \citep{burrows2000} -- it is instead
radiated back to space. From evolutionary models, one finds that
0.1-1\% of the stellar flux impinging on the planet's atmosphere is
sufficient to explain the radius anomaly at the current age of the
planets if it can penetrate to the deep convective zone of the planets
\citep{guillot2002}.
  % FFFFFFFFFFFFF
\begin{figure}[t]
\centering
\includegraphics[width=1.\textwidth]{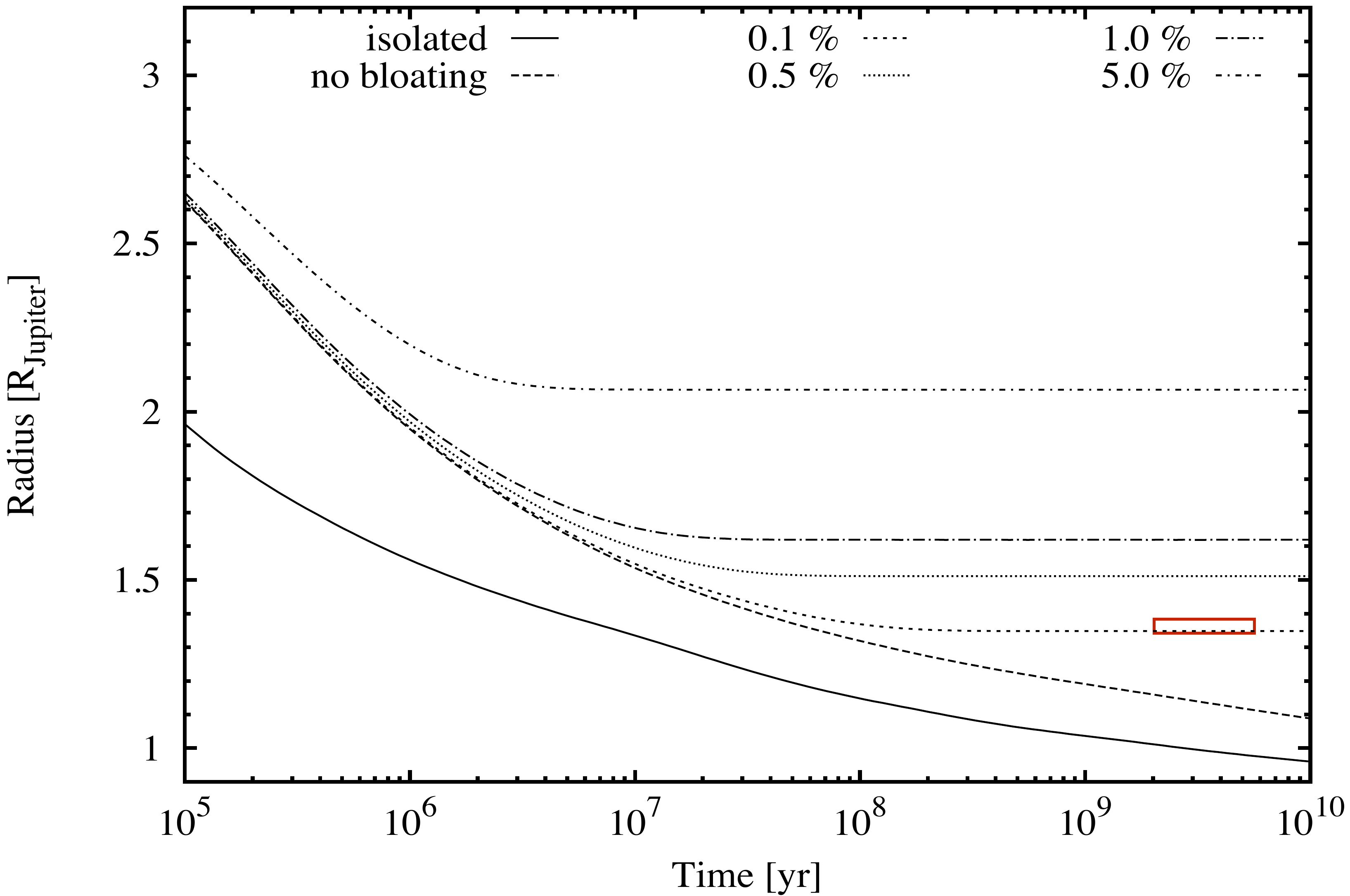}
\caption{Time evolution of the radius of a HD 209458 b-like
  planet. Results are shown for an isolated planet without stellar
  irradiation, an irradiated planet without inflation, and for four
  irradiated planet models where a fraction between 0.1 and 5\% of the
  incoming stellar flux is added to the planet's luminosity in the
  deep convective zone. The red box indicates the observed radius and
  age of HD 209458 b.}
\label{fig:bloating}
\end{figure}
% FFFFFFFFFFFFF
  
This is illustrated in Figure~\ref{fig:bloating}, which displays the
time evolution of the radius of a planet with parameters similar to HD
209458 b ($M=0.69\pm 0.02\,M_{\rm J}$, a = 0.0475$\pm0.0006$ AU, age =
$4.0\pm2.0$ Gyr). The results of six models are illustrated in the
figure: an isolated planet (without stellar irradiation), an
irradiated planet for which stellar irradiation does not come as an
additional energy source for the planet (no bloating, that is no
radius inflation), and for four irradiated planets for which a
fraction of the incoming stellar flux is added to the planet's
intrinsic luminosity at the center (thereby inflating the
planet). According to these calculations, which use the planet
evolution model of \citet{Mordasini15review}, stellar irradiation
increases the radius relative to the isolated case at late times by
about 0.13 $R_{\rm J}$, bringing it to 1.13 $R_{\rm J}$ at 4 Gyr,
which is clearly less than the observed radius (marked by the red box
in Figure~\ref{fig:bloating}). These calculations indicate that about
0.1\% of the incoming flux must be added to the planet luminosity in
the deep convective zone in order to reproduce approximately the
observed radius (for comparison, \citealp{guillot2002} found about
0.08\%).

We note that a difference in radius of about 0.25 $R_{\rm J}$ between
the standard irradiated model and the aforementioned inflated model
corresponds to a much hotter thermodynamics state of the planet as the
partially degenerate equation of state governing the planet's interior
means that the radius is only weakly dependent on temperature. This is
visible from the result that the irradiated planet without additional
energy sources contracted to the observed radius when the planet was
almost 100 times younger than the actual age of the system. The 0.1\%
fraction of the incoming stellar irradiation necessary to bring the
planet to its observed radius corresponds to a luminosity that is
about 90 times bigger than the present-day luminosity of a planet
undergoing standard cooling without inflation.

There are two main mechanisms that can turn stellar irradiation into
an energy flux reaching the deep interior of planets, but note that
none of them can so far explain the radius anomaly of all extrasolar
planets {\it quantitatively} \citep{baraffechabrier2014}:
\begin{enumerate}
\item[-] The first one is the so-called ohmic heating mechanism, for
  which ionized alkali atoms in the winds caused by stellar
  irradiation of the planet produce an electric current that is
  dissipated in the deep layers of the planet
  \citep{batyginstevenson2010,PernaMenou2010}. However, the study by
  \citet{HuangCumming2012a}, who calculated the heating rate more
  self-consistently than earlier works, indicates that ohmic
  dissipation might not be strong enough to explain the radius anomaly
  of the very massive and very hot Jupiters.
\item[-] The second mechanism is the so-called {\it weather noise}
  mechanism \citep{showmanguillot2002}. The strong winds resulting
  from the large temperature contrasts between the day- and
  night-sides of tidally locked planets are found in the numerical
  simulations of atmospheric circulation of \citet{showmanguillot2002}
  to lead to a significant downward transport of kinetic energy. This
  energy is then dissipated in the deep layers, the details of which
  are currently not well understood. Also, the feasibility of a
  sufficiently large downward transport of kinetic energy depends on
  the details of the employed atmospheric circulation model and is
  subject to significant uncertainties \citep{baraffechabrier2014}.
  Lastly, the combination of tides together with the temperature
  contrast between the planetary day- and night-sides may produce work
  in the planet which could again result in an energy flux to the
  planet's deep layers \citep[thermal tides,
  see][]{ArrasSocrates2010}.
\end{enumerate}

Other mechanisms have been proposed to explain the observed radius
anomaly of hot Jupiters, which we briefly outline below:
\begin{enumerate}
\item[-] An increased metallicity in the planet's atmosphere which
  could lead to a slower contraction, because the opacity is expected
  to increase along with the metallicity of the planet and therefore
  to slow down its cooling \citep{burrows2007},
\item[-] A decrease in the efficiency of convective energy transport
  in the deep interior of the planet due to compositional gradients
  and the associated onset of semi-convection, which would also slow
  down the planets' cooling \citep{ChabrierBaraffe2007},
\item[-] Tidal dissipation in the planets' interior resulting from
  star-planet tidal interactions
  \citep[e.g.,][]{BodenheimerLin2001}. While tidal circularization of
  a giant planet (whose eccentricity could arise from interactions
  with unseen planetary companions) could provide significant heating
  in the early evolution of the planet, it is unlikely to remain a
  significant source of radius inflation at the typical ages where
  inflated hot Jupiters are observed
  \citep{lecontechabrier2010a}. Also, tidal dissipation does not
  account for the observed dependency between radius anomaly and
  stellar irradiation flux.
\end{enumerate}

In conclusion of this paragraph, the correlation between the
occurrence of inflated hot Jupiters and stellar irradiation is well
established observationally, but the mechanisms behind it remain
poorly understood. Theorists have yet to come up with a convincing
theory which can explain the observations and quantitatively work when
coupled to planetary structure models.

%==============
\section{Summary}\label{sec:summary}
%==============
The main points to take away from this chapter are the following:
\begin{itemize}
\item[$\bullet$] The planetesimal accretion scenario of planetary
  growth is dictated by self-gravity, and proceeds through runaway
  growth and oligarchic growth. It can account for terrestrial planet
  formation in the inner Solar system. Growth becomes inefficient in
  the oligarchic phase, and growth timescales become exceedingly long
  toward large orbital separations.
\item[$\bullet$] Accretion of small pebbles is assisted by gas drag,
  and is most efficient for marginally coupled particles with stopping
  time to orbital period ratios between roughly 0.1 and 1. Once a
  substantial fraction of solid mass resides in such pebbles, embryos
  that have grown across the transition mass accrete most pebbles
  entering their Hill sphere, and growth timescales are well within
  the lifetime of protoplanetary disks even at large orbital
  separations.
\item[$\bullet$] Both planetesimal and pebble accretions likely play
  important roles in the formation of the Solar system as well as for
  exoplanets, and they may dominate at different phases and in
  different regions of protoplanetary disks.  Global models of planet
  formation have begun to take both accretion scenarios into
  account. Pinning down the uncertainties requires better
  understandings of disk physics.
\item[$\bullet$] The gravitational interaction between planets and
  their protoplanetary disk plays a prominent role in the early
  orbital evolution of planetary systems. Planet-disk interactions
  usually damp the eccentricity and inclination, but can lead to
  inward or outward migration depending on the planet's mass and,
  crucially, the disk's physical properties near the planet (radial
  profiles of the disk's density and temperature, magnetic field,
  turbulence). While tremendous progress has been made in the past few
  years in understanding the physics of planetary migration,
  especially that of low-mass planets, predictive scenarios of planet
  migration require more knowledge of protoplanetary disks in regions
  of planet formation. Pursuing detailed observations of
  protoplanetary disks as well as comprehensive models of the dynamics
  of such disks will help theories of disk migration become more
  predictive.
\item[$\bullet$] While disk migration is certainly inevitable, it is
  one among many mechanisms that shape the architecture of planetary
  systems. The great diversity of extrasolar planetary systems shows
  that disk and high-eccentricity migrations, interactions with host
  and nearby stars all play some role.  (Mis-)aligned hot Jupiters and
  the many multiple super-Earths systems illustrate the plurality of
  mechanisms shaping planetary systems.
\item[$\bullet$] Evolutionary models establish the link between
  formation and observations.  They are keys to estimate the mass of
  young directly imaged planets for which only the age and luminosity
  are known, and to learn about the interior composition and structure
  of planets when only their mass and radius are known like for
  close-in low-mass planets. For this class of planets, the bulk
  composition like the H/He mass fraction and the ice mass fraction in
  the core represent important constraints for their formation
  mode(s). Atmospheric compositional measurements (using transit
  spectra) could help break degeneracies between composition and
  structure in the future, and help constrain the planet's formation
  mechanism and migration history.
\item[$\bullet$] The transition between giant planets and brown dwarfs
  is smooth. Planets with a mass of $\sim 13 M_{\rm J}$ can burn
  deuterium in their interior, just like brown dwarfs. After a rapid
  phase of deuterium burning, a non deuterium-burning planet of 12
  M$_{\rm J}$ and a planet that has depleted its deuterium (say, with
  a mass of 14 M$_{\rm J}$) will have very similar structures,
  comparable to that of low-mass brown dwarfs after their deuterium is
  depleted (except, for a solid core which can, however, dissolve into
  the H/He in time).
\item[$\bullet$] Like for models of planet migration, predictions of
  evolutionary models are tied to the formation process (e.g., hot or
  cold start models), especially for young planets. While many
  exoplanets are well described by the mass-radius or the mass-mean
  density distributions predicted by core accretion linked to
  evolutionary models, there are several ways such models can be
  improved, in particular regarding the planet's internal structure
  (e.g., the occurrence of semi-convection), its material properties
  (via updated equations of states at high pressures and
  temperatures), and models for special evolutionary effects
  like atmospheric escape or radius inflation.
\end{itemize}

% AAAAAAAAAAAAAAAA
\begin{acknowledgements}
  We thank the organizers of the ISSI workshop "The Disk in Relation
  to the Formation of Planets and their Protoatmospheres", which was
  held in Beijing in August 2014. We thank D. Lin, N. Madhusudhan,
  Z. Sandor and S. Udry for stimulating discussions at the
  workshop. We thank Gabriel Marleau for interesting discussions and
  for providing Figure~\ref{fig:ini_cond}, and David Swoboda and Yann
  Alibert for providing the starting data for
  Figure~\ref{fig:mrho}. We also thank Aur\'elien Crida and Bertram
  Bitsch for detailed comments on an earlier draft of this paper, and
  the referee for a detailed and constructive report. XNB acknowledges
  support from Hubble and ITC Fellowships. CM acknowledges the support
  from the Swiss National Science Foundation under grant
  BSSGI0$\_$155816 ``PlanetsInTime''.
 \end{acknowledgements}
% AAAAAAAAAAAAAAAA

\begin{multicols}{2}
\bibliographystyle{ppvi}
%\bibliography{clement,planet,migration_ppvi,chapter4,bloating}

\end{multicols}

\end{document}